\documentclass[sigconf]{acmart}

\newcommand*{\affaddr}[1]{#1} % No op here. Customize it for different styles.
\newcommand*{\affmark}[1][*]{\textsuperscript{#1}}

\usepackage{caption}
\usepackage{amsmath,booktabs}
\usepackage{subcaption}
\usepackage{amsmath,amsfonts}
\usepackage{algorithmic}
\usepackage{textcomp}
\usepackage{xcolor}
\usepackage{smartdiagram}
\usepackage{booktabs}
\usepackage{balance}
\usepackage{flushend}
\usepackage{bm}
\usepackage{listings}
\usepackage{natbib}

\usepackage{xspace}
\usepackage{color}
\usepackage{enumitem}

\usepackage{cleveref}

\usepackage{array,multirow,graphicx}

\usepackage{framed}
\newenvironment{result}{\begin{framed}\centering\it}{\end{framed}}

\newcommand{\recheck}[1]{\textcolor{black}{#1}}
\newcommand{\revise}[1]{\textcolor{black}{#1}}

\newcommand{\approach}{\textsc{GraphCode2Vec}\xspace}

\def\figref#1{Figure~\ref{fig:#1}}
\def\figlabel#1{\label{fig:#1}\label{p:#1}}

\newcounter{notecounter}

\newcommand{\enoteson}{\long\gdef\enote##1##2{{
\stepcounter{notecounter}
{\large\bf \hspace{1cm}\arabic{notecounter} $<<<$ ##1: ##2 $>>>$\hspace{1cm}}}}}
\enoteson

\usepackage{listings}

\usepackage{color}

\definecolor{dkgreen}{rgb}{0,0.6,0}
\definecolor{gray}{rgb}{0.5,0.5,0.5}
\definecolor{mauve}{rgb}{0.58,0,0.82}

\renewcommand{\shortauthors}{Ma and Zhao, et al.}

\begin{document}

% \title{\approach: Generic Code Embedding via \\
% Lexical and Program Dependence Analyses}
% \todo{change title}}

\title{\approach: Generic Code Embedding via \\
Lexical and Program Dependence Analyses}

% Learning Generic Code Representation \\ Using Graph Neural Networks
% Lexical and Semantic Features
% }
% On the Use of Graph Neural Networks in forming Generic Code Representation 

 \author{
 Wei Ma\affmark[1]\affmark[*], Mengjie Zhao\affmark[2]\affmark[*], Ezekiel Soremekun \affmark[1], Qiang Hu\affmark[1], Jie Zhang\affmark[3], Mike Papadakis\affmark[1], Maxime Cordy\affmark[1], Xiaofei Xie\affmark[4], Yves Le Traon\affmark[1]\\
 \affaddr{\affmark[1]SnT, Luxembourg},
 \affaddr{\affmark[2]LMU Munich},
 \affaddr{\affmark[3]University College London}\\
 \affaddr{\affmark[4]Nanyang Technological University}
 }
 \thanks{\affmark[*]\text{these authors contributed equally},\affmark[1]{first\_name.last\_name}@uni.lu, \affmark[2]mzhao@cis.lmu.de, \affmark[3]jie.zhang@ucl.ac.uk, \affmark[4]{xfxie@ntu.edu.sg}}

 \renewcommand{\shortauthors}{Ma and Zhao, et al.}

\begin{abstract}
Code embedding is a keystone in the application of machine learning on several Software Engineering (SE) tasks. To effectively support a plethora of SE tasks, the embedding needs to capture program syntax and semantics in a way that is \textit{generic}. To this end, we propose the \textit{first 
self-supervised pre-training} approach (called \approach) which produces task-agnostic embedding of lexical and program dependence features. \approach achieves this via a  synergistic combination of \textit{code analysis} and \textit{Graph Neural Networks}. \approach is \textit{generic}, it \textit{allows pre-training}, and it is \textit{applicable to several SE downstream tasks}. We evaluate the effectiveness of \approach on four (4) tasks (method name prediction, solution classification, mutation testing and overfitted patch classification), and compare it with four (4) similarly \textit{generic} code embedding baselines (Code2Seq, Code2Vec, CodeBERT, GraphCodeBERT) and \recheck{7} \textit{task-specific}, learning-based methods. In particular, \approach is more effective than both generic and task-specific learning-based baselines. It is also complementary and comparable to GraphCodeBERT (a larger and more complex model). We also demonstrate through a probing and ablation study that \approach learns lexical and program dependence features and that self-supervised pre-training improves effectiveness. 
\end{abstract}

\maketitle
%\keywords{code embedding, code representation, code analysis}

% \todo{replace soundness with a better word, soundness is strange for a heuristic or learning based approach}

% \todo{replace ``task-specific learning-based approaches'' with ``task-specific learning-based approaches''}

%\todo{Provide a motivating example form the results/datasets}

%\input{introduction}

%\todo{address the difference between this work and ProgramGraph~\cite{allamanis2017learning}}

% \todo{are the task-specific approaches code embedding ones? if so why we  do not  use these embedding in the other tasks?}

% \todo{I am not sure we can claim soundness here since the approaches are heuristic by nature. I think we should use different term.}

\begin{figure*}[!bt]
	{\centering
	\vspace{-0.5cm}
	\caption{Motivating example %using Code Clone Detection Task 
	showing (a) an original method (\texttt{LowerBound}), and two behaviorally equivalent clones of the original method, namely (b) a renamed method (\texttt{findLowerBound}), 
% 	with method and variable renaming, 
    and (c) a refactored method (\texttt{getLowerBound}). 
    % with variable/method renaming, and 
% 	as well as 
% re-ordering of method parameters and instructions (\textit{e.g.}, variable initialization and \texttt{if-else} branch). 
% We also examine a direct copy of the original program (a), albeit with method renamed to \texttt{searchLowerBound} (not shown here). %}
% 	{ Notice: \textit{searchLowerBound} is exactly same with \textit{lowerBound} so we do not show it here.
	}
	\vspace{-0.1cm}
	\includegraphics[width=0.95\textwidth]{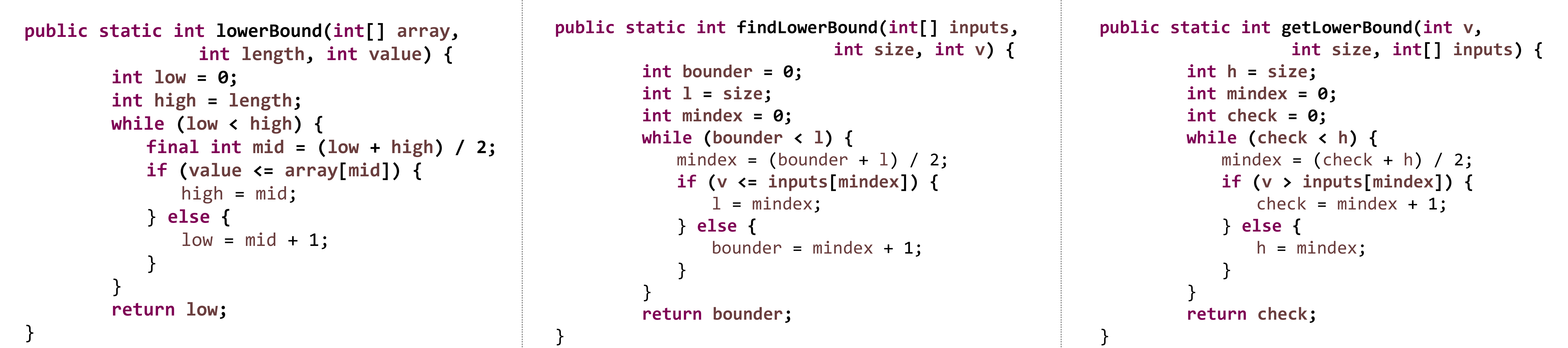}
	\label{fig:example} 
	}\\ 
	\quad \quad\quad \quad \quad \quad (a) Original Method \quad \quad \quad \quad \quad \quad \quad \quad \quad \quad (b) Renamed Method \quad \quad\quad \quad \quad \quad  \quad \quad (c) Refactored Method \quad \quad \quad \quad \quad \quad
 	\vspace{-0.4cm}
\end{figure*}

% \todo{rename paper title and approach/ name} \\
% \todo{if we remove code clone detection, remember to remove all references to task in paper}

\section{Introduction}
\label{sec:introduction}

Applying machine learning to address software engineering (SE) problems often requires a vector representation of the program code, especially for deep learning systems. %~\todo{cite}. 
A na{\"i}ve representation, used in many SE applications, is one-hot encoding that represents every feature with a dedicated binary variable (a vector including binary values)~\citep{seger2018investigation}. 
% \revise{
However, this type of embedding is usually 
% leads to 
a high-dimensional sparse vector because the size of vocabulary is very large in practice, which results in the notorious \textit{curse of dimensionality} problem
% \footnote{\url{https://en.wikipedia.org/wiki/Curse_of_dimensionality}} 
~\citep{bellman2015adaptive}. %Another drawback of 
Besides, one-hot encoding has %is that it has 
\textit{out-of-vocabulary} (OOV) problem, which 
decreases model generalization capability such that it cannot handle new type of data~\citep{urolagin2011generalization}.
% }
%that do not fit well to the majority of neural network toolkits []. 

%
To deal with these issues, researchers use %the use of 
dense and reasonably %\revise{low}\com{we should avoid use low} 
concise %dimensional 
vectors 
% is generally advisable 
to encode program features for specific SE tasks, since  
% as 
they generalise better~\citep{jiang2007deckard, 7582748, deepsim, wang2020detecting}. 
% \revise{
% To deal with this issue and extract good vector representation, deep learning models are widely used to encode code feature for some specific SE tasks \citep{jiang2007deckard, 7582748, deepsim, wang2020detecting} in a decade. 
% Since Programming languages obey the 
% naturalness hypothesis \citep{allamanis2018survey, hindle2016naturalness}, 
% r
More recently, researchers apply natural language processing (NLP) techniques %are leveraged 
to learn the universal code embedding vector for general SE tasks~\citep{feng2020codebert, buratti2020exploring, kanade2020learning, allamanis2017learning, wang2020learning, puri2021project, alon2019code2seq, alon2019code2vec, bui2021infercode, 10.1145/3377811.3380361, ben2018neural, venkatakeerthy2020ir2vec, cummins2020deep, peng2021could, guo2020graphcodebert}. 
% , i.e. code embedding. 
% 
% In essence 
The resulting code embedding represents a mapping from the ``program space'' to the ``latent space'' that captures the different code-used semantics, i.e., the semantic similarities between program snippets. %Key to this
The aim is that similar programs % snippets 
should have similar representations in the latent space. 

% \revise{
% However, most 
% The current 
State-of-the-art code embedding approaches focus either on \textit{syntactic features} (\textit{i.e.}, tokens/AST), or on \textit{semantic features} (\textit{i.e.}, program dependencies) ignoring the importance of combining both features together. For example, 
Code2Vec~\citep{alon2019code2vec} and CodeBERT~\citep{feng2020codebert}) focus on syntactic features, while PROGRAML~\citep{cummins2020deep} and NCC \citep{ben2018neural}) focus on program semantics. 
% features, while  ones. 
%However,\textit{ both features are vital to address several SE tasks},  %This is particularly important because 
%since programs are made up of both features. 
%Besides, several popular SE tasks, such as code clone detection and code classification, require both program semantics and syntax to be effectively addressed. %embedded in the vector representation %of the program 
% to be effective. 
There are few studies using both program semantics and syntax, e.g., GraphCodeBERT~\citep{guo2020graphcodebert}. However, these approaches are not \textit{precise}, they do not obtain or embed the entire program dependence graph. Instead, they estimate program dependence via string matching (instead of static program analysis), then  
% semantic of data and control dependencce by leveraging program static analysis. These models add 
augment AST trees with sequential data flow edges.

\revise{
To address these challenges, we propose the \textit{first approach (called \approach) to synergistically capture syntactic and semantic program features with \textit{Graph Neural Network (GNN)} via self-supervised pretraining}.
%\approach produces generic, task-agnostic embeddings that effectively  capture both syntax and program semantics via combined \textit{lexical embedding} and \textit{dependence embedding},  respectively. 
}
The \textit{key idea} of our approach is to use \textit{static program analysis} and \textit{graph neural networks} to effectively represent programs in the latent space. This is achieved by combining lexical and program dependence analysis embeddings. During lexical embedding, \approach embeds the syntactic features in the latent space via \textit{tokenization}. In addition, it performs dependence embedding to capture program semantics via static program analysis, it derives the program dependence graph (PDG) and represent it in the latent space using Graph Neural Networks (GNN). It then concatenates both lexical embedding and dependence embedding in the program's vector space. This allows \approach to be effective and %ly 
applicable on several downstream tasks. 
% representation of both 
% This is called  \textit{dependencce embedding}. 
% }

To demonstrate the importance of semantic embedding, we compare the similarity of three pairs of programs using our approach, in comparison to a syntax-only embedding approach -- CodeBERT, and GraphCodeBERT, which embeds both syntax and semantic, albeit without program dependence analysis. Consider the example of three program clones in \Cref{fig:example}. This example includes three behaviorally or semantically equivalent programs, that %the \textit{renamed} and \textit{refactored} programs (\Cref{fig:example}(b) and (c)) are clones of the \textit{original} program -- \texttt{lowerBound} (\Cref{fig:example}(a)). 
have low syntactic similarity (i.e., different tokens), but with similar semantic features, i.e.,  program dependence graphs (PDGs). %\Cref{fig:example}(b) is a copy of the ``original'' program (\Cref{fig:example}(a)) called a ``renamed'' program, albeit with only method and variable renaming. Meanwhile, \Cref{fig:example}(c) shows another copy of the ``original'' program (\Cref{fig:example}(a)) but with variable/method re-naming as well as instruction and condition re-ordering, this program is called a ``refactored'' program. Finally, we also compare the original program to a direct copy of it, 
%called ``searchLowerBound'', albeit with only method name re-naming. 
To measure the similarity distance in the latent space, in addition to the example code clones (\Cref{fig:example}), we randomly select 10 other different code methods (from GitHub) 
% and also include one direct clone of \textit{lowerBound} ( renamed \textit{searchLowerBound}) 
without any change to establish a baseline for comparing all approaches. \revise{To this end, we compute the average cosine similarity distance for all 91 
program pairs ($\frac{14 \times 13}{2}$) for reference to show that all approaches report similar scores for all randomly selected 91 pairs (\Cref{tab:example_sim}).\footnote{\revise{The purpose of computing the average cosine similarity of all 91 code pairs is to establish a meaningful reference for comparing embeddings and to serve as a sanity check. We expect the mean of the cosine similarity of a set of randomly selected pairs of code clones and non-clones to lie around zero for all approaches (range -1 to 1).}}}
%of methods 
% as a %the 
% comparison 
% } 
For all three approaches, the similarity between the ``original program'' and a direct copy of the program with only method name renaming to ``searchLowerBound'', 
% the ``renamed program'' (syntax only changes) 
is well captured with an almost perfect cosine similarity score for all approaches ($1$ or $0.99$). 
% (\textit{see} \Cref{tab:example_sim}). 
Likewise, the cosine similarity of the original program and the ``renamed'' program (\texttt{findLowerBound}) is mostly well captured by all approaches, since they all embed program syntax, albeit with lower cosine similarity scores for CodeBERT (0.61) and GraphCodeBERT (0.70), in comparison to our approach (0.99). 

Meanwhile, CodeBERT fails to capture the semantic similarity %be
% Notably, we show that a syntax-only embedding fails to capture the semantic  similarity 
between the ``original program'' and the ``refactored program'' (\texttt{getLowerBound}), even though they are behaviorally similar and share similar program dependence. This is evidenced by the low cosine similarity score (0.51),  
% This is due to the fact that 
because it does not account for semantic information in its embedding, especially the similar program dependence graph shared by both programs. Lastly, GraphCodeBERT performs slightly better than CodeBERT (0.70 vs. 0.51), but lower than our approach (0.99). This is due to lack of actual static program analysis in the embedding of GraphCodeBERT, since it only applies a heuristic (string matching) to estimate program dependence, it is \textit{imprecise}. This example demonstrates the importance and necessity of embedding precise dependence information. 

\begin{table}[!t]
  \caption{Cosine Similarity of three behaviorally/semantically similar  program pairs from our motivating example, using %about Motivation Example Code for 
  GraphCodeBERT, CodeBERT and \approach 
%   about the original implementation \textit{lowerBound} with the clone versions
}
  \vspace{-0.35cm}
 \scalebox{0.7}{
\begin{tabular}{|l|c|c|c|} %l|} 
\hline
 \multirow{2}{*}{\textbf{Program Pairs}} & \textbf{Graph-} &  \multirow{2}{*}{\textbf{CodeBERT}} & \multirow{2}{*}{\textbf{\approach}} \\ % & \textbf{Approach} \\  
 & \textbf{CodeBERT} &   &   \\ \hline % & \textbf{Average} \\ 
\texttt{searchLowerBound} \& \texttt{lowerBound} & 1 & 0.99 & 1 \\
\texttt{findLowerBound} \& \texttt{lowerBound} &  0.70 & 0.61 & 0.99 \\
\texttt{getLowerBound} \& \texttt{lowerBound} & 0.70 & 0.51 & 0.99 \\
\hline
Average of 91 pairs &  -0.05 & -0.06 & -0.03 \\
\hline
\end{tabular}
}
\vspace{-1.6em}
\label{tab:example_sim}
\end{table}

% \begin{table}[!t]
%   \caption{Cosine Similarity about Motivation Example Code for GraphCodeBERT, CodeBERT and \approach about the original implementation \textit{lowerBound} with the clone versions}
%   \vspace{-0.35cm}
%  \scalebox{0.7}{
% \begin{tabular}{|l|c|c|c|c|c|} %l|} 
% \hline
%  \multirow{2}{*}{\textbf{Method Name}} & \textbf{Graph-} &  \multirow{2}{*}{\textbf{CodeBERT}} &\multicolumn{3}{c|}{\textbf{\approach}} \\ % & \textbf{Approach} \\  
%  & \textbf{CodeBERT} &   & \textbf{semantic} & \textbf{syntactic}  & \textbf{sem+syn}  \\ \hline % & \textbf{Average} \\ 
% searchLowerBound & 1 & 0.99 & 1 & 1 & 1 \\
% findLowerBound &  0.70 & 0.61 & 0.97 & 0.99 & 0.99 \\
% getLowerBound & 0.70 & 0.51 & 0.98 & 0.99 & 0.99 \\
% \hline
% Average of 91 pairs &  -0.05 & -0.06 & -0.07 & -0.03 & -0.03 \\
% \hline
% \end{tabular}
% }
% \vspace{-1.6em}
% \label{tab:example_sim}
% \end{table}

\revise{A key ingredient of \approach is \textit{self-supervised pretraining}. Even though task-specific learning based approaches (e.g., CNNSentence~\cite{ohashi2019convolutional}) learn the vector representation of code without pre-training, they are non-generic and less effective. Applying their learned vector representation to other (SE) tasks requires re-tuning model parameters, and the lack of pretraining reflects in their performance. % in comparison to pretrained models. 
As an example, our evaluation (in RQ1 \autoref{sec:results}) showed that our self-supervised pretraining approach %stage 
improves effectiveness when compared to \recheck{7} task-specific approaches (i.e., without pretraining) addressing two (SE) tasks (%code clone detection, 
solution classification and patch classification). To further demonstrate the importance of \textit{self-supervised pretraining}, % in \approach, 
we 
%evaluate the impact of pretraining %on the effectiveness of our approach 
%by 
compare the effectiveness of \approach with and without pretraining using two downstream tasks. Overall, we demonstrate that our self-supervised pretraining improves effectiveness by 28\% %more effective 
(\textit{see} RQ3). % and \autoref{tab:config-analysis} in \autoref{sec:results}). 
}
% \revise{
%In our evaluation, we examine the 
% We have evaluated the 
%effectiveness of our approach in embedding both syntactic and semantic features, using three downstream SE tasks and seven (7) subject programs. 
% Notably, w

To evaluate \approach,
%In our evaluation, 
we compare it to four generic code embedding approaches, and \recheck{7} task-specific learning-based applications. %, for the code clone detection and solution classification tasks. 
We also investigate the stability and learning ability of our approach through sensitivity, ablation and probing analyses. 
Overall, we make the following \textit{contributions}: %}
% }

\noindent
\textbf{\textit{\recheck{Task-specific learning-based applications.}}} 
% \revise{
We introduce the 
% \recheck{
automatic application of \approach to solve specific downstream SE tasks, without extensive human intervention to adapt model architecture.
% }
%\recheck{automatic application of \approach to solve specific downstream SE tasks}, without intensive fine-tuning or human intervention. 
In comparison to the state-of-the-art task-specific learning-based approaches (\textit{e.g.}, \revise{ODS~\citep{ye2021automated}}
%ProgramGraph~\citep{allamanis2017learning}
), our approach does %not require any effort to tune the hyperparamet the pre-trained model to be applicable to a downstream task (\Cref{sec:approach}). \approach is the \recheck{first} approach to be easily amenable to downstream tasks without additional fine-tuning or \recheck{human intervention}. %In o
not require any effort to tune the hyper-parameters to be applicable to a downstream task (\Cref{sec:approach}).
Our evaluation on \revise{two downstream tasks, %code clone detection, 
solution classification and patch classification}, showed that \approach %either 
outperforms %or is comparable to 
the state-of-the-art task-specific learning-based applications: \revise{For 
% most (\todo{X out of X}) 
% both
all tasks it outperforms all
task-specific applications  
% , and it is comparable to the best-performing task-specific approach for (\todo{X out of X}) tasks 
(RQ1 in \Cref{sec:results}).}
%\com{we fine-tune the model. I am not sure if no requiring intensive fine-tuning may confuse the reader.}
% \item[Performance improvement via Model Pre-training.] To improve model performance, we propose a variant of \approach for %the application of 
% model pre-training. 
% % for \approach (\autoref{sec:approach}). 
% We evaluate the influence of model pre-training on the performance of \approach for our downstream SE tasks. %We pre-train our model with a large dataset and then fine-tune the model on several downstream tasks. 
% Our evaluation results show that \approach's model pre-training %with \approach 
% achieves significant performance improvements on our downstream tasks (\todo{RQ?} in \autoref{sec:results}). %\approach pre-training 
% It improves performance by \todo{X}\% across all downstream tasks, on average.
% }

% \begin{description}
% \item[
\noindent 
\textbf{\textit{Generic Code embedding}}. 
% \revise{
We propose a novel and generic 
%automatic 
code embedding learning approach (\textit{i.e.,} \approach) that captures the 
lexical, control flow and data flow features of programs through a novel combination of \textit{tokenization}, \textit{static code analysis} and \textit{graph neural networks} (GNNs). \revise{To the best of our knowledge, \approach is the first code embedding approach
to precisely 
capture syntactic and semantic program features with GNNs via self-supervised pretraining.}
% that uses GNNs derived from \emph{precise} program dependencies. % information. % in the vector space and is wi
%\approach is generically applicable to several downstream tasks, making it 
%In our comparative evaluation using three (3) downstream tasks, \recheck{three} program datasets and \recheck{four} (4) generic baselines, 
We demonstrate that \textit{\approach is effective} (RQ2 in \Cref{sec:results}): \revise{\textit{{It outperforms all syntax-only generic code embedding baselines}}}. 
We provide our pre-trained models and generic embedding for public use and scrutiny.\footnote{\url{https://github.com/graphcode2vec/graphcode2vec}} 
% }
% }

% \vspace{-0.42cm}

\noindent
\textbf{\textit{Further Analyses.}} % of \approach.}} 
% \revise{
We extensively evaluate the \textit{stability} and \textit{interpretability} of
% the \textit{soundness} or 
% correctness is still strong. it is a black box actually. What we do is to observe its output to guess it.
our approach by conducting %using a combination of 
\textit{sensitivity}, \textit{probing} and \textit{ablation} analyses. 
% In addition, w
We also investigate the impact of configuration choices 
% in terms of 
(i.e., pre-training strategies and GNN architectures) on the effectiveness of our approach on downstream tasks. Our evaluation results show that \approach %is \textit{sound}, it 
\textit{effectively learns lexical and program dependence features}, it is \textit{stable} and insensitive to the choice of GNN architecture or pre-training strategy (RQ3 in \Cref{sec:results}).\footnote{
% As a reminder, 
In the rest of this work, we interchangeably use the terms ``lexical'' and ''syntactic'' interchangeably, as well as ``(program) dependence'' and ``semantic''. Such that the terms ``lexical embedding'' and ``syntactic embedding'' refer to the embedding of program syntax, and the terms ``dependency embedding'' and ``semantic embedding'' refer to the embedding of program dependence information.
% Because program dependence graph can represent program semantic, semantic embedding is treated as the same to dependence embedding in this work. Similarly, we use ``lexical'' and ``syntactic'' interchangeably throughout the paper.}
% \end{description}
}

% \revise{
%The rest of this paper is organized as follows: In \Cref{sec:background}, we discuss closely related work %and provide background 
%on %the area of 
%code embedding. \Cref{sec:approach} describes %briefly introduces 
%our approach (\approach). %with a motivating example. 
%After describing our experimental setup (\Cref{sec:experimental-setup}),  \Cref{sec:results} discusses our experimental results.
% and future outlook (\autoref{sec:discussions}). 
%Finally, we highlight the limitations and threats to the validity of our approach (\Cref{sec:threats-to-validity}), and we close with conclusion (\Cref{sec:conclusion}). 
% }

%\input{background}

\section{Background}
\label{sec:background}
\subsection{Generic code embedding}
% \revise{
% In this work, \textit{generic code embedding} approaches 
% This refers to 
We discuss methods that learn general-purpose code representations to support several downstream tasks. 
% \textit{i.e.}, t
These approaches are not designed for a specific task.
% , but are applicable to several downstream (SE) tasks. 
There are three major types of generic code embedding 
approaches, % in these area, 
namely \textit{syntax-based}, \textit{semantic-based} and \textit{combined semantic and syntactic} approaches (\textit{see} \Cref{tab:state-of-the-art-details}). 
% \Cref{tab:state-of-the-art-details} presents the state-of-the-art generic code embedding approaches.
% Researchers have proposed several generic code embedding approaches, including approaches employing neural models for representing code (snippets), \textit{e.g.}, via code vector (\textit{e.g.}, Code2Vec~\citep{alon2019code2vec}), machine translation (\textit{e.g.}, Code2Seq~\citep{alon2019code2seq}) or transformers (\textit{e.g.}, CodeBERT~\citep{feng2020codebert}).
% }

\noindent
\textbf{\textit{Syntax-based Generic Approaches:} }
% Syntactic code embedding 
These approaches 
% Researchers have proposed several code embedding approaches that 
encode program snippets, either by dividing the program into \textit{strings}, 
lexicalizing %lexing is not a word
them into \textit{tokens} or parsing the program into a\textit{ parse tree or abstract syntax tree (AST)}. Syntax-only generic embedding approaches include Code2Vec~\citep{alon2019code2vec}, Code2Seq~\citep{alon2019code2seq},  CodeBERT~\citep{feng2020codebert}, C-BERT~\citep{buratti2020exploring}, InferCode\citep{bui2021infercode}, CC2Vec~\citep{10.1145/3377811.3380361}, AST-based NN~\citep{zhang2019novel} and ProgHeteroGraph~\citep{wang2020learning} (\textit{see} \Cref{tab:state-of-the-art-details}). Notably, these approaches use neural models for representing code (snippets), \textit{e.g.}, via code vector (\textit{e.g.}, Code2Vec~\citep{alon2019code2vec}), machine translation (\textit{e.g.}, Code2Seq~\citep{alon2019code2seq}) or transformers (\textit{e.g.}, CodeBERT~\citep{feng2020codebert}). 
Code2Vec~\citep{alon2019code2vec} is an AST-based code representation learning model %trained 
that represents code snippets as single fixed-length code vector. %by decomposing 
It decomposes a program into a collection of paths using an %in its 
AST and learns %ing 
the atomic representation of each path while simultaneously learning how to aggregate the set of paths. 
% The authors demonstrated on a large, cross-project corpus that it is effective for method name prediction and it outperforms previous techniques. % by more than 75\%. 
Code2Seq~\citep{alon2019code2seq} is an alternative code embedding approach that uses Sequence-to-sequence (seq2seq) models, adopted from neural machine translation (NMT), to encode code snippets. 
% It leverages the syntactic structure of programming languages to encode source code by representing code snippets as the set of paths in the program's AST, then uses attention to select the relevant paths while decoding. 
% The authors show that it significantly outperforms previous models on two tasks, namely method name prediction and  code captioning. 
% \revise{
CodeBERT~\citep{feng2020codebert} is a bimodal pre-trained model for programming language (PL) and natural language (NL) tasks,  which uses transformer-based neural architecture to encode code snippets. Besides, CodeBERT \citep{feng2020codebert}, C-BERT~\citep{buratti2020exploring} and Cu-BERT \citep{kanade2020learning}  are BERT-inspired approaches, these methods adopt similar methodologies to learn code representations as BERT~\cite{devlin2018bert}.
% Its evaluation via fine-tuning on two NL-PL tasks (natural language code search and code documentation generation) shows that CodeBERT achieves state-of-the-art performance. 
% Meanwhile,  C-BERT~\cite{buratti2020exploring} pre-trains a large transformer-model on raw source code, then tests if the model can discover AST features.  Cu-BERT~\citep{kanade2020learning} is similar to C-BERT, but if focuses on embedding programs written in the \texttt{Python} programming language.
% }

% \revise{
% Our approach (
\approach is similar to the aforementioned generic code embedding methods, it is also a general-purpose code embedding approach that captures syntax by 
lexicalizing %lexing 
the program into tokens (\textit{see} \Cref{tab:state-of-the-art-details}). However, all of the aforementioned generic approaches are syntax-based, % (i.e., based on ASTs, strings or tokens), 
none of these approaches account for program semantics (i.e., data and control flow). Unlike these approaches, \approach additionally captures %also accounts for 
program semantics %, it extracts and encodes program dependencies 
% (\textit{i.e.}s, data and control flow information) 
via static analysis. %, then encodes the extracted program dependence features. 
% of the program. 
% In addition, it also accounts for lexical features of the program by employing lexical analysis to encode code (snippets). Our evaluation on several programming tasks shows that it outperforms the state-of-the-art. % on downstream SE tasks. 
In this paper, we compare our approach (\approach) to the three (3) most popular and recent syntax-based generic code embedding approaches, namely Code2Vec~\citep{alon2019code2vec}, Code2Seq~\citep{alon2019code2seq} and CodeBERT~\citep{feng2020codebert} (\textit{see} %RQ1 in 
\cref{sec:results}). 
% }

\noindent
\textbf{\textit{Semantic-based Generic Approaches:}}
% \revise{
% Semantic-based approaches 
This refers to code embedding methods that capture \textit{only} semantic information such as control and data flow dependencies in the program. \textit{Semantic-only generic approaches} include NCC~\citep{ben2018neural} and PROGRAML~\citep{cummins2020deep}. On one hand, NCC~\citep{ben2018neural} extracts the contextual flow graph of a program by building an LLVM intermediate representation (IR) of the program. It then applies word2vec~\citep{mikolov2013distributed} to learn code representations. 
% In its evaluation, the authors demonstrated that NCC (even without fine-tuning) outperforms specialized approaches for performance prediction. 
On the other hand, PROGRAML~\citep{cummins2020deep} is a language-independent, portable representation of whole-program semantics for deep learning, which is designed for data flow analysis in compiler optimization. It adopts message passing neural networks (MPNN)~\citep{messagepassing} to learn LLVM IR representations. 
% The authors evaluation of PROGRAML showed that it learns standard data flow information and results in improved performance on downstream compiler optimization tasks. 
In contrast to these approaches, \approach captures both semantics and syntax. %ctic information. 
% }

% \revise{
% Similar to \approach, t
\noindent 
\revise{
\textbf{\textit{Combined Semantic and Syntactic -based Approaches:}}} 
% Besides, t
There are generic approaches that capture both syntactic and semantic features such as IR2Vec~\citep{ben2018neural}, OSCAR~\citep{peng2021could}, ProgramGraph~\citep{allamanis2017learning}, ProjectCodeNet~\cite{puri2021project} and GraphCodeBERT~\citep{guo2020graphcodebert}. IR2Vec~\citep{ben2018neural} and OSCAR~\citep{peng2021could} use LLVM IR representation of a program to capture program semantics. 
% In particular, IR2Vec~\citep{venkatakeerthy2020ir2vec} learns LLVM IR hierarchical code representation and OSCAR~\citep{peng2021could} learns LLVM IR representations 
% using a transformer architecture.
Meanwhile,  ProgramGraph~\citep{allamanis2017learning} uses GNN to learn syntactic and semantic representations of code from ASTs augmented with data and control edges. ProgHeteroGraph leverages abstract syntax description language (ASDL) grammar to learn code representations via heterogeneous graphs~\citep{wang2020learning}. 
%\enote{mzhao}{why do we need this sentence here? i think we can just discard it\recheck{ProjectCodeNet~\cite{puri2021project} provides quite useful source code datasets and some general code representation for SE tasks.}
%ProjectCodeNet~\cite{puri2021project} employs a  multi-layer perceptron (MLP) with bag of tokens to encode programs
% . In particular, its MLP encode program snippets 
%as token-frequency vector that stores operator and keyword tokens. }
% It also provides several configurations including CNN configuration with token sequence, 
% % which uses the same token set with the MLP but treats them as a sequence with order. In addition, it provide four 
% and GNN configuration encodings  (GIN, GIN-V, GCN, and GCN-V) with Simplified Parse Tree (SPT)~\citep{parr2013definitive}. 
% , which can be applied with or without  additional virtual nodes (-V) that are
% connected to all the nodes in the graph.
Finally, GraphCodeBERT~\cite{guo2020graphcodebert} is built upon CodeBERT~\cite{feng2020codebert}, but in addition to capturing syntactic features it also accounts for semantics by employing data flow information in the pre-training stage. 
%, by augmenting sequences of the data flow with the AST Tree. 
% }

% \revise{
Similar to these approaches, our approach (\approach) learns both syntactic and semantic features. In this work, we compare \approach to GraphCodeBERT because it is the most recent state-of-the-art and closely related approach to ours, %similar to 
% our approach, 
since it captures both syntax %tic 
and semantics %program features 
(\textit{see} RQ2 \cref{sec:results}).
% }
% \todo{discuss solution classification task-specific learning-based approaches}
% \todo{discuss ProjectCodeNET}

\begin{table}[!bt]
\vspace{-0.5em}
  \caption{%\todo{add approaches for mutation testing and patch overfitting}
  Details of the state-of-the-art Code Embedding approaches. ``Semantic'' or ``Sem'' means program dependence, % (control and/or data flow), 
  and ``Syntactic'' or ``Syntax'' refers to strings, tokens, parse tree or AST-tree. Symbol ``\checkmark'' means the approach supports a feature, and ``$\times$'' means it does not support the feature. 
%   \todo{add the three solution classification baselines in table and background discussion}
%   \todo{DeepSim \citep{deepsim}}
 }
 \vspace{-0.4cm}
 \scalebox{0.7}{
\begin{tabular}{|l|l|l|l|l|l|l|l|} %l|} %l|l|l|l|l|}
\hline
\multicolumn{2}{|c|}{\multirow{2}{*}{\textbf{Type}}} & \multirow{2}{*}{\textbf{Approaches}} %\textbf{Code Embedding} %& \textbf{Generic (\checkmark) or } 
& \multirow{2}{*}{\textbf{Syntactic}} & \multirow{2}{*}{\textbf{Semantic}}  %& \textbf{Pre-training} & \textbf{Source Code (\checkmark)}  
&  \multicolumn{2}{c|}{\textbf{Granularity}}  %& \textbf{Intensive} & \textbf{X} & \textbf{X}  
\\ %\hline
\multicolumn{2}{|c|}{}  %& \textbf{Task-specific ($\times$)} 
& & & & \textbf{Method} & \textbf{Class} 
%& \textbf{fine-tuning (\checkmark)} & \textbf{X} & \textbf{X} 
\\ \hline
% \parbox[t]{2mm}{\multirow{3}{*}{\rotatebox[origin=c]{90}{Task-\\specific}}}
\multirow[c]{7}{*}{\rotatebox[origin=c]{90}{\parbox{2cm}{\textbf{Task-specific}}}} & 
\multirow[c]{2}{*}{\rotatebox[origin=c]{90}{\parbox{1cm}{\textit{Syntax}}}}
% & DECKARD~\citep{jiang2007deckard} & \checkmark & $\times$ & $\times$ &  \checkmark \\ %\hline
% % \rotatebox{90}{\parbox{2mm}{\multirow{4{*}{yourtexthere}}} 
%  %\hline
% & & RtvNN~\citep{7582748} & \checkmark  & $\times$ & \checkmark  & \checkmark \\ %\hline
% & 
& CNNSentence~\cite{ohashi2019convolutional} & \checkmark &  $\times$ &  $\times$ &  \checkmark \\
& & OneCNNLayer~\cite{pinter2018classification} & \checkmark &  $\times$ &  $\times$ &  \checkmark \\
& & SequentialCNN~\cite{gilda2017source} & \checkmark &  $\times$ &  $\times$ &  \checkmark \\
\cline{2-7}
& \multirow[c]{4}{*}{\rotatebox[origin=c]{90}{\parbox{0.7cm}{\textit{Both}}}} & 
% DeepSim~\citep{deepsim} & \checkmark  & \checkmark  & \checkmark  &  $\times$ \\
% & & FA-AST~\citep{wang2020detecting} & \checkmark &  \checkmark &  \checkmark &  \checkmark \\ 
% & & 
\revise{SimFeatures}~\cite{wang2020automated} & \checkmark & \checkmark & $\times$ & \checkmark \\
& & \revise{Prophet}~\cite{long2016automatic} & \checkmark & \checkmark & $\times$ & \checkmark \\
& & \revise{PatchSim}~\cite{xiong2018identifying} & \checkmark & \checkmark & $\times$ & \checkmark \\
& & \revise{ODS}~\cite{ye2021automated} & \checkmark & \checkmark  & $\times$ & \checkmark \\
 %\hline
% & SPT~\citep{parr2013definitive} & & & & \\ 
\hline
% & & & & & \\ \hline
\multirow[c]{15}{*}{\rotatebox[origin=c]{90}{\parbox{2cm}{\textbf{Generic}}}} & 
\multirow[c]{8}{*}{\rotatebox[origin=c]{90}{\parbox{1.8cm}{\textit{Syntax-only}}}} &  CodeBERT~\citep{feng2020codebert}  & \checkmark &  $\times$ & \checkmark &  $\times$ \\

& & Code2Vec~\citep{alon2019code2vec} & \checkmark & $\times$  &  \checkmark & $\times$ \\ 
& & Code2Seq~\citep{alon2019code2seq} & \checkmark & $\times$ & \checkmark & $\times$ \\ 
& & C-BERT~\cite{buratti2020exploring} & \checkmark & $\times$ & \checkmark & \checkmark  \\ 
& & InferCode~\citep{bui2021infercode} & \checkmark  & $\times$  &  \checkmark &  \checkmark \\ 
& & CC2Vec~\citep{10.1145/3377811.3380361} & \checkmark & $\times$ & \checkmark &  \checkmark \\ 
& &  AST-based NN~\citep{zhang2019novel} & \checkmark & $\times$ & $\times$ & \checkmark \\ 
& & ProgHeteroGraph~\citep{wang2020learning} &  \checkmark & $\times$ & \checkmark &   $\times$  \\ 
\cline{2-7}
& 
\multirow[c]{2}{*}{\rotatebox[origin=c]{90}{\parbox{0.7cm}{\textit{Sem.}}}} 
& NCC~\citep{ben2018neural} & $\times$ &  \checkmark & \checkmark & \checkmark  \\ 
& & PROGRAML~\citep{cummins2020deep} & $\times$ & \checkmark & \checkmark &  \checkmark \\ 
\cline{2-7}
& \multirow[c]{6}{*}{\rotatebox[origin=c]{90}{\parbox{1cm}{\textit{Both}}}}  & IR2Vec~\citep{ben2018neural} & \checkmark & \checkmark &  \checkmark & \checkmark \\
& & OSCAR~\citep{peng2021could} & \checkmark & \checkmark  &  \checkmark &  \checkmark  \\
& & ProgramGraph~\citep{allamanis2017learning} & \checkmark &  \checkmark &  \checkmark & \checkmark  \\ 
& & ProjectCodeNet~\citep{puri2021project} & \checkmark &  \checkmark &  $\times$ &  \checkmark \\
& & GraphCodeBERT~\citep{guo2020graphcodebert} & \checkmark & \checkmark & \checkmark &  $\times$  \\ 
\cline{3-7}
& & \textbf{\approach} %(this paper) 
 & \checkmark & \checkmark &  \checkmark & \checkmark \\
\hline
\end{tabular}
}
\vspace{-0.65cm}
\label{tab:state-of-the-art-details}
\end{table}

% \todo{mention vulnerability detection approaches from ICSE r}
\vspace{-0.2cm}
\subsection{\recheck{Task-specific learning-based applications}}%Not sure these approaches generate code embedding? 
% \revise{
% In the following, we 
% Let us discuss %all the 
% task-specific learning-based methods, \textit{i.e.},  
% specialised ML-based techniques that are developed %for a 
% to address specific (SE) downstream task. 
% , called . 
% task-specific ML-based approaches employed in our experiments. 
% Researchers have proposed numerous specialised % task-specific 
% learning-based approaches %. % have been proposed 
% to solve several downstream SE tasks. %For instance, 
% Notably, r
Researchers have proposed specialised learning-based techniques to tackle specific (SE) downstream tasks, e.g.. %the challenges of 
% \textit{code clone detection}~\cite{jiang2007deckard, wang2020detecting,7582748} 
\textit{patch classification}~\cite{ye2021automated, long2016automatic} and 
\textit{solution classification}~\cite{ohashi2019convolutional,pinter2018classification,gilda2017source}. In our experiments, we consider specialised learning approaches for both tasks.  % the specific SE task of
% code clone detection and solution classification. 
This is 
% and solution classification tasks because they have 
because these tasks have several software engineering applications, especially during software maintenance and evolution~\cite{ohashi2019convolutional, ye2021automated, long2016automatic}. %jiang2007deckard, sheneamer2016survey,7582748}. 
% These specialized learning-based %Task-specific 
% approaches %for code clone detection 
% can be classified into \textit{syntax-based} and \textit{semantic-based} task-specific approaches. %All of t
% The evaluated task-specific approaches 
% % The syntax-based code embedding approaches 
% employed in this paper for code clone detection includes DECKARD~\citep{jiang2007deckard}, RtvNN\citep{7582748}, DeepSim~\citep{deepsim} and FA-AST~\citep{wang2020detecting}, 
% % 
% For solution classification, we employed 
% % ProjectCodeNet~\cite{puri2021project} as a baseline. 
% CNNSentence~\cite{ohashi2019convolutional}, OneCNNLayer~\cite{pinter2018classification} and 
% SequentialCNN~\cite{gilda2017source} as specialised learning baselines. 
\Cref{tab:state-of-the-art-details} %provides details on 
highlights 
% the 
details %properties 
% of all seven (7) 
of 
our task-specific learning methods. % for code clone detection. 
\noindent
\textbf{\textit{Solution classification:}}
% For solution classification, w
Let us 
describe the state-of-the-art learning-based approaches for solution classification. Most of these approaches are syntax-based and adopt convolution neural networks (CNNs) to classify programming tasks. 
% solved by a program, using the source code. 
SequentialCNN~\cite{gilda2017source} applies a CNN to predict the language/tasks from code snippets using lexicalized tokens represented as a matrix of word embeddings. 
% It was evaluated 
% against five other state-of-the-art learning-based approaches using thousands of source code files from GitHub and the top 10 most used programming languages on GitHub. The authors demonstrated that it is more effective than the state-of-the-art learning-based approaches.  
CNNSentence~\cite{ohashi2019convolutional} is similar to SequentialCNN
%~\cite{gilda2017source} we dont need to cite at each occurrence
since it also uses CNNs, except that it classifies source code without relying on keywords, \textit{e.g.}, variable and function names. It instead considers the structural features of the program in terms of 
tokens that characterize the process of arithmetic processing, loop processing, and conditional branch processing. 
% Its evaluation on fundamental programming tasks (\textit{e.g.}, graph theory, computational theory, computer graphics, artificial intelligence, and discrete mathematics) showed that it predicts the correct task category with high accuracy. 
Finally, OneCNNLayer~\cite{pinter2018classification} also uses CNN for solution classification. It firstly pre-processes the program to remove unwanted entities (\textit{e.g.}, comments, spaces, tabs and new lines), then tokenizes the program to generate the code embedding %, %vector space 
using word2vec. The resulting embedding %their approach 
% the program
includes the token connections and their underlying meaning in the vector space. 
% Its evaluation on several tasks and three programming languages (Java, C\# and Python) showed that it has a high accuracy in predicting the task addressed by the program. %source code. 
% }

% \revise{
% \noindent
% \textbf{\textit{Mutant Classification:}}
% \todo{x}
% }

\noindent
\revise{
\textbf{\textit{Patch Classification:}} % of Over-fitting Patches:}}
% We discuss the state of the art (learning-based) patch classification techniques, especially
These are techniques designed to determine the correctness of patches (i.e., identify correct, wrong or over-fitting patches). %these techiques
These learning-based techniques can be % are typically % including %hese includes 
static (e.g., ODS~\cite{ye2021automated}), dynamic (e.g., Prophet~\cite{long2016automatic}), heuristic-based (e.g., PatchSim~\cite{xiong2018identifying}) or hybrid (e.g., SimFeatures~\cite{wang2020automated}). 
% such as ODS~\cite{ye2021automated}, PatchSim~\cite{xiong2018identifying}, Prophet~\cite{long2016automatic} and SimFeatures~\cite{wang2020automated}. 
\Cref{tab:state-of-the-art-details} provides details of these approaches. Notably, they all capture both syntactic information (e.g. via AST) and program dependence information (e.g., via execution paths or control flow information). 
For instance, PatchSim~\cite{xiong2018identifying} is a \textit{heuristic approach} that leverages the behavioral similarity of test case executions to determine patch correctness by leveraging %, it tracks via a 
the complete path spectrum of test executions.  
% \citet{long2016automatic} proposed Prophet, a \textit{dynamic patch classification approach} that ranks correct patches by learning a probabilistic model of code from correct human patches, it  also captures program dependencies %semantics by employing 
% via a schema that tracks the insertion of control flow statements in patches. %SimFeatures~
Meanwhile, \citet{wang2020automated} proposed (SimFeatures --) a \textit{hybrid strategy that identifies correct patches by integrating static code features with dynamic features or (test) heuristics}. %-based approaches}. 
SimFeatures combines %achieves this by combining 
a learned static code model with %other 
dynamic or heuristic-based information (such as %program dependencies by tracking 
the dependency similarity between a buggy program and a patch) 
% patch classification approaches via 
using majority voting. 
% The authors demonstrated that this combination improves over the performance of existing patch classification techniques. 
More recently, \citet{ye2021automated} proposed a supervised learning approach (called ODS) that employs static code features of patched and buggy programs to determine patch correctness, specifically to classify over-fitting patches. It uses supervised learning on extracted static code at the AST level to learn a probabilistic model for determining patch correctness. ODS also tracks program dependencies by tracking control flow statements. %In this paper, 
For this task, we compare %\aproach to %the performance of 
\approach to ODS, PatchSim, Prophet and SimFeatures (\textit{see} \Cref{sec:results}). %, since ODS has been demonstrated to outperform all of the aforementioned approaches~\cite{ye2021automated}. 
% \revise{ODS}~\cite{ye2021automated} uses three categories of features, i.e., code description features, repair pattern features and contextual syntactic features, constructed based on the expert experience. PatchSim~\cite{xiong2018identifying} considers the dynamic test execution as the auxiliary of the patch over-fitting detection. PatchSim\cite{xiong2018identifying} utilizes the longest common subsequence(LCS) as the distance measurement to embed the code into a scalar value.
}

In this work, we compare %our approach (
\approach %) 
to the aforementioned \revise{seven (7) %four (4) 
learning-based methods %approaches 
for 
%code clone detection, 
solution classification 
and patch classification }% of over-fitting patches }%. %among other tasks 
(\textit{see} %RQ2 in 
\Cref{sec:results}). 
\section{Approach}
\label{sec:approach}
\begin{figure}[t]
    \vspace{-0.3cm}
    % \vspace{-0.5em}
	\centering
	\caption{Overview of \approach 
	}
	\includegraphics[width=0.5\textwidth]{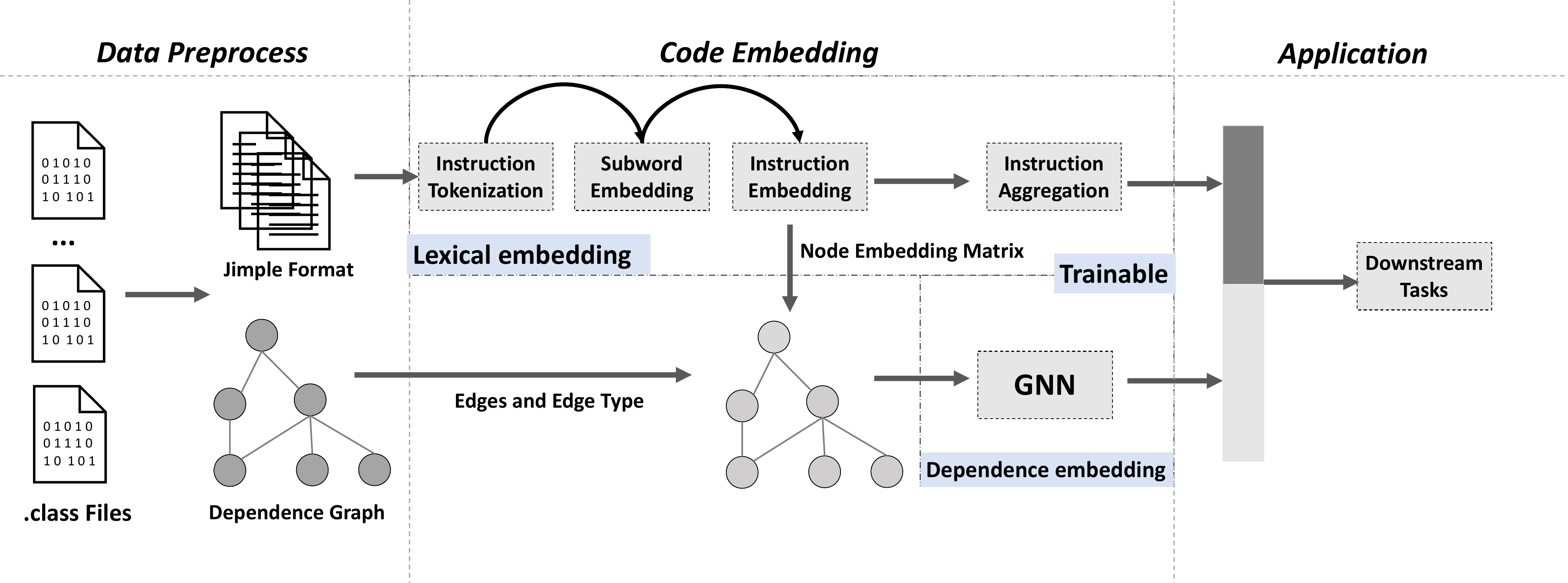}
	\figlabel{fig:overview}
	\vspace{-1.23cm}
\end{figure}

\subsection{Overview}
\figref{fig:overview} illustrates the steps and components of our approach.
%In our data pre-processing step, our approach takes as input a Java program, convert it into the Jimple intermediate representation and obtain its program dependence graph (PDG).
First, \approach takes as input a Java program (i.e. a set of class files) that is converted to a Jimple intermediate representation. Secondly, \approach employs Soot~\citep{vallee2010soot} to obtain the program dependence graph (PDG) by feeding the class files as input. %Then, to embed a program, 
From the resulting Jimple representation and PDG, %program dependence graphs, 
\approach learns two program %ming intermediate-representation instruction-level 
embeddings, namely a lexical embedding and a dependence embedding. These two embeddings are ultimately concatenated to form the final code  embedding.
% we perform lexical embedding and dependencce embedding, then concatenate both embeddings to obtain the overall program representation. 

To achieve \textit{lexical embedding}, our approach first tokenizes the Jimple instructions obtained from our pre-processing step into sub-words. Next, given the sub-words, our approach learns sub-word embedding using word2vec~\citep{word2vecpaper}. Then, it %our approach 
learns the instruction embedding by representing every Jimple instruction as a sequence of subwords embeddings using a bi-directional LSTM (BiLSTM, \Cref{sec:lstm}). The forward and backward hidden states of this BiLSTM %\revise{is used} 
allows to build the instruction embeddings. \approach employs a BiLSTM since it learns context better: BiLSTM can learn both past and future information while LSTM only learns past information. %, which means BiLSTM . 
Finally, it aggregates multiple instruction embeddings using element-wise addition, in order to obtain the overall lexical program embedding. % of the .

% % to obtain the 
% % \revise{
% Our lexical embedding represents Jimple instruction separately. We tokenize each Jimple instruction into subwords. We, then, learn a subword embedding using word2vec
% \citep{word2vecpaper}. To embed instructions, we feed the embedding of the constituent subwords into a bidirectional LSTM (BiLSTM, \Cref{sec:lstm}). The forward and backward hidden states of this BiLSTM \revise{is used} to build the instruction embeddings. 
% %\com{Provide a reason for that.}  
% \revise{BiLSTM can learn both of past and future information while LSTM only learns from the past information, which means BiLSTM can learn the context better than LSTM.}
% Finally, we use element-wise addition to aggregate multiple instruction embeddings and get the program lexical embedding.

To learn the \textit{dependence embedding}, \approach applies % create 
a Graph Neural Network (GNN) \citep{scarselli2008graph} to embed Jimple instructions %as well as 
and their dependencies. Each node in the graph corresponds to a Jimple instruction and contains the (dependence) embedding of this instruction. Node attributes are from lexical embeddings.
%Before training, each node is initialized to the lexical embedding of the corresponding instruction.
%not true, not initialized!! it is computed fly-way!!
The edges of the graph represent the dependencies between instructions. 
%\com{Please explain why, also how do you get the data flows and are they precise? } 
Our approach %We 
considers 
% \revise{four types of 
the following program dependencies: data flow, control flow and method call graphs. \approach uses intra-procedural analysis \citep{10.1145/24039.24041} to extract data-flow and control-flow dependencies by invoking Soot \citep{vallee2010soot}. %, a static analysis tool that is able to extract dependencce graphs from Jimple. 
Then, it builds method call graphs via class hierarchy analysis \citep{10.5555/646153.679523}. %for the low-cost call graph construction.
% }

The training of GNNs is an iterative process where, at each iteration, the embedding of each node $n$ is updated based on the embedding of the neighboring nodes (i.e., nodes connected to $n$) and the type of $n$'s edges \citep{xu2018powerful, zhou2020graph}. The \emph{message passing function} determines how to combine the embedding of the neighbors -- also based on the edge types -- and how to update the embedding $n$ based on its current embedding and the combined neighbors' embedding. The dependence embedding of an instruction is %, then, 
the embedding of the corresponding node at the end of the training process.

Finally, after obtaining lexical embedding and dependence embedding, our approach concatenates both embeddings to obtain the overall program representation. 
\vspace{-1em}
\subsection{Lexical embedding}
% \subsubsection{
\noindent
\textbf{\textit{Step 1 - Jimple code tokenization:}} 
% \label{sec:tokenizationproc}
The first crucial step of \approach is to properly tokenize Jimple code into meaningful ``tokens'', %for which we can then 
to learn the vector representations. 
The traditional way to tokenize code is to split it on whitespaces. However, this manner is inappropriate for two reasons. First, whitespace-based tokenization often results in long tokens such as long method names (e.g., ``getFunctionalInterfaceMethodSignature''). Long sequences often have a low frequency in a given corpus, which subsequently leads to an embedding of inferior quality. Second, whitespace-based tokenization is not able to process new words that do not occur in the training data -- these out-of-vocabulary words are typically replaced by a dedicated ``unknown'' token. This is an obvious disadvantage for our approach, whose goal is to support practitioners to analyze diverse programs -- which may then include words that did not occur in the programs used to learn the embedding.

To address this challenge, we
% We, instead, 
tokenize the Jimple code into \emph{subwords} \citep{bpepaper,wordpiecepaper,sentencepiecepaper}, which are units shorter than words, e.g., morphemes. Subwords have been widely adopted in representation learning systems for texts \citep{bpemb,mbpepaper,bert,t5paper} as they solve the problem of overly long tokens and out-of-vocabulary words. 
% \revise{
New code programs can be smoothly handled using short tokens representation, by limiting the amount of long, but different tokens. Subwords get rid of the almost-infinite character combinations that are common in 
% we usually meet when we consider 
many program codes. %} 
For example, this is the reason why BERT uses wordpiece subwords \cite{wordpiecepaper}, and XLNet \cite{xlnetpaper} and T5 \cite{t5paper} use sentence-piece subwords. Similarly, \approach uses sentence-piece subwords. % in this work. 
When using subwords, the long token ``getFunctionalInterfaceMethodSignature'' is split into ``get'', ``Functional'', ``Interface'', ``Method'' and ``Signature''. It is worth noting that most of the subwords are in fact words, e.g., ``get'' \citep{jurafsky}.

% \subsubsection{
\noindent
\textbf{\textit{Step 2 - Subword embedding with word2vec:}} 
% \label{sec:word2vec}
Given a subword-tokenized Jimple code corpus $\mathcal{C}$ with vocabulary size $|\mathcal{C}|$, our approach learns a subword embedding matrix $\mathbf{E} \in \mathbb{R}^{|\mathcal{C}| \times d}$ where $d$
% \revise{
is a hyperparameter referring to the embedding dimension ($d$ is usually set to 100). It uses the popular Skip-gram with negative sampling (SGNS) method in word2vec \cite{word2vecpaper} to produce $\mathbf{E}$. %Concretely, two matrices $\mathbf{O}, \mathbf{E} \in \mathbb{R}^{|\mathcal{C}| \times d}$ are randomly initialized. For each subword $w_t$ in $\mathcal{C}$, its embedding vector from $\mathbf{E}$ and $\mathbf{O}$ are denoted as $\mathbf{w}_{t,\mathbf{E}} \in \mathbb{R}^d$ and $\mathbf{w}_{t,\mathbf{O}} \in \mathbb{R}^d$ respectively.  The training objective is to maximize the following log likelihood:
%\begin{equation}
%\sum\limits_{w_p \in \mathcal{P}(w_t)}
%log( \sigma (\mathbf{w}_{t,\mathbf{E}}^\intercal \mathbf{w}_{p,\mathbf{O}}))\ 
%-
%\sum\limits_{w_i \in \mathcal{N}(w_t)}
%log( \sigma (\mathbf{w}_{t,\mathbf{E}}^\intercal \mathbf{w}_{i,\mathbf{O}})),
%\eqlabel{sgnsobj}
%\end{equation}
%\noindent where $\mathcal{P}(w_t)$ is the group of context subwords for
%$w_t$, i.e., subwords occurring together with $w_t$ in $\mathcal{C}$;
%$\mathcal{N}(w_t)$ is the group of non-context subwords of $w_t$ sampled
%from the whole corpus. In our experiments, we set the context window size to 8. % $\sigma(\cdot)$ is the sigmoid function
%$\sigma(x) = 1 / (1+e^{-x})$.  
%After training, 
And $\mathbf{E}$ is utilized as the subword embedding matrix \citep{word2vecpaper}. %\com{why? Please explain the rationale behind these choices otherwise it feels unimportant.}\recheck{this choice is pretty ad-hoc and is what the word2vec authors are doing so i add the citation}

% As a sanity check, we manually verified that SGNS produces embeddings such that related program tokens have  representations with high similarity in the embedding space. For example, \tabref{nearestexamples} shows the nearest neighbors of three querying words ``Double'', ``get'', and ``Throw''. It can be easily observed that the subword embeddings indeed capture token semantics in the code data. For example, the nearest neighbors of ``Double'' are mostly the data types in Java; operation ``get'' has other operations like ``create'' and ``toString'' as its neighbors. 
%\com{That is nice, though it remains unclear whether the subwording is good.}\recheck{the reason is explained in the previous comment modification.}

% \begin{table}[t]
% \caption{Subword embedding examples: nearest neighbors.}
% \centering\small
% \vspace{-1em}
% \begin{tabular}{c|c|c}
% Double & get      & Throw      \\ \hline
% Float  & add      & Finally    \\
% Long   & create   & Catch  \\
% Short  & toString & Else     \\
% double & new     & Implement
% \end{tabular}
% \tablabel{nearestexamples}
% \vspace{-2.5em}
% \end{table}

% \subsubsection{
\noindent
\textbf{\textit{Step 3 - Instruction embedding:}} 
\label{sec:lstm}
After forming the subword embeddings, \approach represents every Jimple instruction as a sequence of subword embeddings $(\mathbf{w}_0, \mathbf{w}_1, ..., \mathbf{w}_n)$, by using a bidirectional LSTM (BiLSTM). The role of BiLSTM is to learn the embedding of the instruction from the subword sequence of the instruction.
%\com{Not sure what aggregate means here.}
%\todo{@Wei: can't we give more details about the LSTM parameters? Also justify why a bidirectional LSTM is appropriate here. + (for me) refer to the pre-training section #Wei: the settign are given in the pretraing in Experiment setting up-section.} 
Let $\overrightarrow{\mathbf{h}_t}$ and $\overleftarrow{\mathbf{h}_t}$ be the forward hidden state and backward hidden state of LSTM after feeding the final subword. Then, it forms the instruction embedding by concatenating $\overrightarrow{\mathbf{h}_t}$ and $\overleftarrow{\mathbf{h}_t}$, denoted as $\mathbf{x}=(\overrightarrow{\mathbf{h}_t}, \overleftarrow{\mathbf{h}_t})$.

% \subsubsection{
\noindent
\textbf{\textit{Step 4 - Instruction embedding aggregation:}} 
\label{sec:token_aggregator} 
The last step in the process of forming lexical embedding is the aggregation of the instruction embeddings in order to form the overall program lexical embedding. The reason why we aggregate instruction-level embedding as opposed to learning an embedding for the whole program is that LSTMs work with sequences of limited length and thus,  truncate the instructions into small sequences (not exceeding the maximal length). After tokenization, a program can have many subwords and if one directly consider all subwords in the program, one needs to cut these subwords into the limited sequence length for LSTM and result in information loss. 
%\com{Unclear, I am not sure how subwords fit in here since we talk about instruction sequences.}

Our approach uses element-wise addition as the token aggregation function. This operation allows the aggregation of multiple instruction embeddings while keeping a limited vector length. 

\vspace{-1em}
\subsection{Dependence embedding}
% \subsubsection{
\noindent
\textbf{\textit{Step 1 - Building method graphs:}} A method graph is a tuple $G=(V, E, \mathbf{X}, \mathbf{K})$, where $V$ is the set of nodes (i.e. Jimple instructions), $E$ is the set of edges (dependence relations between the instructions), $\mathbf{X}$ is the node embedding matrix (which contains the embedding of the instructions) and $\mathbf{K}$ is the edge attribute matrix (which encodes the dependencies that exist between instructions). 
%To initialize $X$, our approach uses the lexical instruction embeddings output by the LSTM, as described above. 
% Again never use initialize here!
For each node $n$ there is a column vector $\mathbf{x_n}$ in $\mathbf{X}$ such that  $\mathbf{x_n}=(\overrightarrow{\mathbf{h}}_t, \overleftarrow{\mathbf{h}}_t)$ (instruction embedding).

To define $E$ and $\mathbf{K}$, 
% \revise{
our approach extracts data-flow and control-flow dependencies by invoking Soot \citep{vallee2010soot, 10.1145/24039.24041}. %} 
Then, \approach introduces an edge between two nodes if and only if the two corresponding instructions share some dependence.  

% \subsubsection{
\noindent
\textbf{\textit{Step 2 - Building program graphs: }}
A program graph consists of 
a pair $\mathcal{P}=( \mathcal{G} , \mathcal{R})$ where $\mathcal{G}=\{G_0, G_1, ... , G_m\}$ is a set of method graphs and where $\mathcal{R} \subseteq \mathcal{G}^2$ is the
call relation between the methods, that is, 
$(G_i, G_j) \in \mathcal{R}$ if and only if the method that $G_i$ represents calls the method that $G_j$ represents. To represent this relation in the GNN, \approach introduces an entry node and an exit node for each method and edges linking those nodes with caller instructions.

% \subsubsection{
\noindent
\textbf{\textit{Step 3 - Message passing function: }}
% \label{sec:gnn_component}
The exact definition of the message passing function depends on the used GNN architecture. 
% \revise{
We choose the widely-used GNN  architectures with linear complexity \citep{wu2020comprehensive} that has been successfully applied in various application domains. 
% }
%\todo{Justify the choice of these architectures} 
\approach employs % consider 
four GNN architectures, namely Graph Convolutional Network (GCN; \citet{kipf2016semi}), GraphSAGE \citep{hamilton2017inductive}, Graph Attention Network (GAN; \citet{velivckovic2017graph}), Graph Isomorphism Network (GIN; \citet{xu2018powerful}).
% , and Variational Graph Auto-Encoder (VGAE; \citet{kipf2016variational}). GCN defines a graph convolutional operator on the Laplacian matrix as the message passing function, similar to the convolutional operator in convolutional neural networks \citet{lenetpaper}. GraphSAGE \citep{hamilton2017inductive}
% directly aggregates the neighbors' embedding (e.g. it computes their mean) instead of using Laplacian
% matrix. GIN uses element-wise sum. GAN uses
% a self-attention mechanism to assign different weight values to
% the neighbors during updating a node representation. VGAE uses the
% encoder-decoder architecture to get the node representation. It
% assumes that the node representation follows a Gaussian distribution
% so that the encoder learns parameters of the distribution. Then, the
% decoder reconstructs the graph from the output of the encoder.

% \subsubsection{
\noindent
\textbf{\textit{Step 4 - Learning the dependence embedding:}}
The dependence embedding of each instruction is obtained by running the message passing function on all nodes for a pre-defined number of iterations, %\revise{
i.e., the number of GNN layers.
%\todo{@Wei: not sure what is your stop condition -- is it a predefined number of iteration or a fixed point? Might depend on the GNN architecture, I guess. Please elaborate. Answer: itertaions means the number of layers you use GNN, it is hierarchical.}. 
Once these instruction embeddings have been produced, \approach aggregates them using the global attention pool operation \citep{li2015gated} in order to produce the program-level dependence embedding. %\revise{
Attention mechanism can make program-level dependence embedding consider more important nodes (instructions). %} 
%\todo{@Wei, please justify why you use this operation}. 
%\revise{In our experimental settings, this program embedding consists of an 600-sized vector.} %\todo{@Wei, please replace XXX by the vector size (the output of what you previously named the READOUT function). It should in the experiments part.}.

The dependence embeddings that GNN produces depend on the learnable parameters of (a) the message passing function and (b) bidirectional LSTM. These parameters can be automatically set to optimize the effectiveness of \approach either directly on the downstream task or on some pre-training objectives, as described hereafter.

In the end, 
our approach uses a concatenation operator to get the program embedding vector. Concatenation has been shown to be an effective method to fuse features without information loss when using DNN \citep{ghannay2016word, 7298594,
  szegedy2016rethinking, larsson2016fractalnet, huang2017densely,
  oyedotun2020deep}. Although the dependence embedding inherently encodes the lexical embedding, the importance of lexical inherently fades away as the semantic representation is learnt. Our ablation study (see RQ3 in \Cref{sec:results}) later reveals the benefits of concatenating an explicit lexical embedding with the dependence embedding.
  
\subsection{Pre-training}
\label{sec:pretraining}
% \revise{
Self-supervised learning has been applied with success for pre-training deep learning models \cite{erhan2010does, liu2019roberta, bertology}. %Self-supervised learning 
It allows a model %is an intermediate form of supervised learning and unsupervised learning where a model 
to learn how 
to perform 
% (often, classification or regression) 
tasks %from data 
without human supervision~\citep{AndrewZisserman, mundhenk2018improvements} %, the goal is to 
by learning a universal embedding that can be fine-tuned to solve multiple downstream tasks. In this work, we 
% }
% \revise{ 
% implemented 
employed three (3) self-supervised learning strategies %we combine 
to pre-train the BiLSTM and GNN in \approach, namely \textit{node classification}, \textit{context prediction}~\citep{hu2019strategies}, and \textit{variational graph encoding} (VGAE) \citep{kipf2016variational}. 
% In the following we procide details of 
% We discuss each pre-training strategy. 
% 
Node (or Instruction) classification trains the model to infer the type of an instruction, given its embedding. Context prediction requires the model to predict a masked node representation, given its surrounding context. Variational graph encoding (VGAE) learns to encode and decode the code dependence graph structure. Note that these pretraining procedures \emph{do not} require any human-labeled datasets. The model learns from the raw datasets without any human supervision. 

\section{Experimental Setup}
\label{sec:experimental-setup}

%In this section, we describe the evaluation setup for our experiments evaluating the utility of %and  results 
% for 
%our approach (\textit{i.e.}, \approach). 

\smallskip\noindent
\textbf{Research Questions: } 
% \todo{Motivate RQs and explain their implications...}
% \revise{
Our research questions (RQs) are designed to evaluate 
% We evaluate 
the \textit{effectiveness} %and \recheck{\textit{soundness}} 
of \approach. In particular, we compare the \textit{effectiveness} of \approach to %that of 
%'s code embedding in comparison to 
the state-of-the-art in \textit{task-specific} and \textit{generic} code embedding methods (\textit{see} RQ1 and RQ2). This is to demonstrate the utility of \approach in solving downstream tasks, in comparison to specialised learning-based approaches tailored towards solving specific SE tasks (RQ1) and other general-purpose code embedding approaches (RQ1). 
% in code embedding. 
We also examine if %the \recheck{\textit{soundness}} of 
\approach %in 
effectively embeds lexical and program dependence features in the latent space, and how this impacts its effectiveness on downstream tasks (\textit{see} RQ3). 
The first goal of RQ3 is to demonstrate the validity of our approach, i.e., analyse that it indeed embeds lexical and dependence features as intended via \textit{probing analysis}. In addition, we analyse the contribution of lexical embedding and dependence embedding to its effectiveness on downstream tasks by conducting an \textit{ablation study}. We also investigate the \textit{sensitivity} of our approach to the choices in \approach's  framework, e.g., \textit{model pre-training (strategy)} and \textit{GNN configuration}. These experiments allow to evaluate the influence of these choices on the effectiveness of \approach. % on downstream tasks.
% }

% \revise{
Specifically, we ask the following research questions (RQs):
% }

% \begin{itemize}[wide, labelwidth=!, labelindent=0pt]
% \item 
%[\textbf{RQ1}] 

% \item 
\noindent
\textbf{RQ1 Task-specific learning-based applications:} 
% \revise{
Is our approach (\approach) effective in comparison to the state-of-the-art \textit{task-specific} learning-based applications? \revise{What is the benefit of capturing semantic features in our code embedding?}

\noindent
\textbf{RQ2 Generic Code embedding:} 
% \revise{
How effective is our approach (\approach), %code embedding, 
in comparison to the state-of-the-art syntax-only generic code embedding approaches? \revise{What is the impact of capturing both syntactic and semantic features (i.e., program dependencies) in code embedding? How does \approach compare to GraphCodeBERT, a larger and more complex model?}

\noindent
\textbf{RQ3 Further Analyses:} 
\revise{
What is the impact of model pre-training on the effectiveness of \approach?}
Does our approach %(\approach) 
effectively capture lexical and program dependence features? What is the contribution of lexical embedding or dependence embedding to the effectiveness of our approach on downstream tasks? Is our approach \textit{sensitive} to the %cWhat is the impact of the 
 choice of GNN? 
%  and pre-training strategy, and how do they %se choices 
%  impact its effectiveness on downstream SE tasks?

\noindent
\textbf{\\Baselines:}  
% \todo{XXXX}
% \revise{
We compare the effectiveness of \approach to several state-of-the-art code embedding approaches (aka \textit{generic baselines}), and specialised or \textit{task-specific learning-based applications}. 
% including generic and task-specific approaches. 
On one hand, \textit{generic baselines} refers to code embedding approaches that are designed to be general-purpose, i.e., they provide a code embedding that is amenable to address several downstream tasks. On the other hand, \textit{task-specific} baselines refers to learning-based approaches that address a specific 
% are targeted at solving a specific 
downstream SE task, e.g., patch classification. %code clone detection. 
\revise{
\autoref{tab:state-of-the-art-details} provides details about these baselines 
% the state-of-the-art generic baselines and task-specific baselines 
for %code clone detection, 
solution classification and patch classification.} % of over-fitting patches.} 
% each baseline code embedding approach. 
Specifically, we evaluated \approach in comparison to four (4) generic code embedding approaches, namely Code2Seq \citep{alon2019code2seq}, Code2Vec \citep{alon2019code2vec}, CodeBERT  \citep{feng2020codebert} and GraphCodeBERT \citep{guo2020graphcodebert} (\textit{see} RQ2 in \autoref{sec:results}). 
% \recheck{
We have selected these generic baselines %were selected 
because they have been evaluated against several well-known state-of-the-art code embedding methods and demonstrated considerable improvement over them. Besides, these approaches are recent, popularly used %(e.g., (Graph)CodeBERT) 
and have been applied on many downstream (SE) tasks. 
% }

\revise{
For task-specific learning-based approaches, we consider solution classification, 
% code clone detection 
and patch classification.
% the classification of over-fitting patches. %tasks 
% because 
% \recheck{
These are popular SE downstream tasks that have been studied %well investigated in the community, especially  
using learning-based approaches. }
% and they have been demonstrated to outperform several approaches for these tasks (\textit{see} \cref{sec:background}}. 
% For code clone detection, we employed four code clone detection approaches, namely DECKARD~\citep{jiang2007deckard}, RtvNN~\citep{7582748}, DeepSim~\citep{deepsim} and FA-AST~\citep{wang2020detecting} (\textit{see} RQ1 in \autoref{sec:results}). %Meanwhile, w
\revise{
We utilised three (3) specialised learning-based baseline for the solution classification task, namely
CNNSentence~\cite{ohashi2019convolutional}, OneCNNLayer~\cite{pinter2018classification} and SequentialCNN~\cite{gilda2017source}. 
% \revise{
% Finally, w
We also used all four patch classifiers (Prophet~\cite{long2016automatic}, PatchSim~\cite{xiong2018identifying}, SimFeatures~\cite{wang2020automated} and ODS~\cite{ye2021automated}).} %ODS~\cite{ye2021automated} as the specialised learning-based baseline for the patch classification task.}
% of classifying over-fitting patches. }
These task-specific baselines have been selected because 
% \recheck{
they have been shown to outperform other proposed learning-based approaches for these tasks. For instance, 
% \recheck{FA-AST~\cite{wang2020detecting} has been shown to outperform four well-known state of the art baselines including ASTNN, DECKARD and RTvNN.}  
% Likewise, 
SequentialCNN~\cite{gilda2017source} has been evaluated against five other learning-based approaches and demonstrated to be more effective. \revise{%Likewise, 
ODS~\cite{ye2021automated} has also been shown to be more effective and efficient than the three other patch classifiers.
% (Prophet, SimFeatures and PatchSim). 
% PathSim~\cite{xiong2018identifying}, a known classification technique to assess the correctness of (APR) patches.
}
%\vspace{-0.5em}

\smallskip\noindent
\textbf{Subject Programs:} 
% \todo{XXXX}
% \revise{
In our experiments, we employed \revise{eight (8) subject programs} written in \texttt{Java}. % programming language. 
\autoref{tab:subject_programs} provides details about each of our subject programs and their experimental usage. Notably, we employ \revise{four (4) publicly available programs for the downstream tasks, namely Defects4J~\cite{just2014defects4j}, Java-Small~\citep{alon2019code2vec}, and Java250~\citep{puri2021project}}. % and \recheck{GCJ~\citep{deepsim}}}. 
These datasets were employed for our comparative evaluation (\textit{see} RQ1 and RQ2). \revise{We chose these datasets because they are popular and have been employed in the evaluation of our downstream tasks in previous studies~\cite{puri2021project, deepsim, alon2019code2seq, ye2021automated}.} Besides, we employed Java-Small and Java250 in our ablation study where we evaluate the contribution of lexical and dependence embedding to the effectiveness of \approach (RQ3). We chose these two datasets for this task because 
% \recheck{
they correspond to tasks that require lexical and semantic information to be effectively addressed.
% } 
% In addition, t
To further analyze \approach (\textit{see} RQ3), we employed the Concurrency dataset~\citep{garcia2017jada, do2005supporting} and collected two (2) subject programs (named LeetCode-10 and M-LeetCode) from LeetCode\footnote{https://leetcode.com/}. 
% For LeetCode-10, we collected 100 solutions from 10 different problems in LeetCode. 
We use these programs to investigate the difference between capturing lexical and  dependence information. In particular, the Concurrency dataset contains different concurrent code types, which have similar syntactic/lexical features but different structure information. We mutated LeetCode-10 to create  M-LeetCode dataset. % called M-LeetCode. 
Our mutation preserves lexical features, but modifies semantic or program dependence features such that LeetCode-10 and M-LeetCode have the same lexical features, but different semantics. % information. 
For example, a simple dependence mutant involves switching outer and inner loops. 
% or moving the statements in a loop outside the loop. 
We utilize 
LeetCode-10, M-LeetCode and Concurrency for the probing analysis of our approach (\approach). 
% }

\begin{table}[t]
  \begin{center}
  \caption{Details of Subject Programs 
%   \todo{add justification in text/discussion}
  } % (\texttt{Java})}
   \vspace{-0.35cm}
  {\footnotesize
  \begin{tabular}{|l|r|l|} %l|} %r|r|}
  \hline
     \textbf{Subject} & \multirow{2}{*}{\textbf{\#Progs.}} & %\textbf{Average} %& \textbf{Average} 
     %& \textbf{Maturity} 
     \multirow{2}{*}{\textbf{Tasks/Analyses}}  \\ %\multirow{2}{*}{\textbf{Justification}} \\ 
          \textbf{Program} &  &  %& %\textbf{(\#Tasks)} 
         %\textbf{\#Methods} & 
         %\textbf{Size (LOC)} & \textbf{(1$^{\text{st}}$ Commit)} & %\textbf{(Pre-training)
         \\ \hline
    % \multirow{2}{*}{
    \texttt{Java-Small} %} 
    & 
    % \multirow{2}{*}{
    11 %} %(101\~K) & 1452\~K &  
    & Method Name Prediction and Ablation Studies \\ % & \todo{XXXX} \\
    % & &  & \\ 
    % \hline
    % \multirow{2}{*}{
    \texttt{Java250} %} 
    & 75000 %(2.3) & 32 &  
    & Solution Classification and Ablation Studies \\ %\hline
    % & & and Ablation Studies & \todo{XXXX} \\ 
    % \hline
    % \texttt{GCJ} & 1665 %(1.1) %& 50 &  
    % & Code Clone Detection \\ %& \todo{XXXX} \\
    \revise{\texttt{Defects4J}}  & \revise{15 \& 5} & \revise{Mutant Prediction and Patch Classification} \\
    % \texttt{(GCJ)\footnote{https://github.com/parasol-aser/deepsim/issues/9, GCJ only contains 1665 projects.}} & & & & \\
    % \hline
    \texttt{LeetCode-10} & 100 %(1) & 20 &  
    & Probing Analysis \\ % & \todo{XXXX} \\ \hline
    \texttt{M-LeetCode} & 100 %(1) & 20 &  
    & Probing Analysis \\ % & \todo{XXXX}  \\ \hline
     \texttt{Concurrency} & 46 %(7.4)  & 78.8 &  
     & Probing Analysis \\ % & \todo{XXXX} \\ \hline
    \texttt{Jimple-Graph} & 1976 %(1762) & 38\~K &  
    & Model Pre-training \\ % & \todo{XXXX} \\ 
    \hline
    %\texttt{Average} & & & & \\ \hline
    \end{tabular}}
  \label{tab:subject_programs}   
 \vspace{-0.6cm}
\end{center}
\end{table}

% \begin{table}[t]
%   \begin{center}
%   \caption{Details of Subject Programs} % (\texttt{Java})}
%   {\footnotesize
%   \begin{tabular}{@{}l|r|r|r|r@{}} %r|r|}
%      \textbf{Subject} & \textbf{\#Programs} & \textbf{\#Methods} & \textbf{Size} & \textbf{Maturity} \\ 
%           \textbf{Program} & \textbf{(\#Tools)} & \textbf{(\#Tasks)} & \textbf{(in KLOC)} & \textbf{(1$^{\text{st}}$ Commit)} \\ \hline
%     \texttt{Java-Small} & 11 & 1109895 & 15969540 & \\ \hline
%     \texttt{Java250} & 75000 & 175046 & 2369713 & \\ \hline
%     \texttt{Google Code } & 1665 & 1767 & 82811 & \\
%     \texttt{Jam (GCJ)\footnote{https://github.com/parasol-aser/deepsim/issues/9, GCJ only contains 1665 projects.}} & & & & \\
%     \hline
%     \texttt{LeetCode-10} & 100 & 100 & 1981 & \\ \hline
%     \texttt{M-LeetCode} & 100 & 100 & 1997 & \\ \hline
%      \texttt{Concurrency} & 46 & 339  & 3627 & \\ \hline
%     \texttt{Jimple-Graph} & 1976 & 3482283 & 75323085 & \\ \hline
%     \texttt{Average} & & & & \\ \hline
%     \end{tabular}}
%   \label{tab:subject_programs}   
% \end{center}
% \end{table}

\smallskip\noindent
\textbf{Downstream Tasks:} 
% \todo{XXXX}
% \revise{
In our evaluation, we considered \revise{four (4) major software engineering tasks, namely, \textit{mutant prediction}, \textit{patch classification}, \textit{method name prediction}, and \textit{solution classification}. % and \textit{code clone detection}.
} These are popular downstream SE tasks that have been investigated in the community for decades. %, using both program analysis and ML techniques. %machine learning. 
For these \revise{four} tasks, we evaluated \approach in comparison to four \textit{generic baselines}, namely Code2Seq~\citep{alon2019code2seq}, Code2Vec~\citep{alon2019code2vec}, CodeBERT~\citep{feng2020codebert} and GraphCodeBERT~\citep{guo2020graphcodebert}. 
% In addition, 
% we employed four other specialised learning-based approaches for code clone detection and three ML approaches for solution classification. 
\Cref{tab:subject_programs} provides details on the subject programs employed for each downstream tasks. 
In the following, we provide further details about the experimental setup for each task evaluated in this paper. 
% }

\smallskip\noindent
\textit{Method Name Prediction:} 
% \revise{
This refers to the task of predicting the method name of a function in a program, given \revise{a set of method names and} the body of the function as inputs~\citep{bui2021infercode}. This task is useful for automatic code completion during programming. 
% For this task, the goal is to predict names that are human-comprehensible and indicative of the role of the methods. Models trained for this task learn to predict useful method names that capture semantic similarities, combinations, and analogies~\citep{alon2019code2vec}. 
In our experiment, all four generic baselines were evaluated %applied 
for this task. %In addition, w
We evaluated this task using the Java-Small dataset, since it was designed %specifically 
for this task 
in previous studies~\cite{alon2019code2vec} (\textit{see} \Cref{tab:subject_programs}).  
% \todo{models and data set used for tasks and why}
% we compared 4 baselines to our approach (\approach), namely, Code2Seq~\citep{alon2019code2seq} , Code2Vec~\citep{alon2019code2vec}, CodeBERT~\citep{feng2020codebert} and GraphCodeBERT~\citep{guo2020graphcodebert}. We also refer to some comparable results from InferCode~\citep{bui2021infercode}.  \autoref{tab:state-of-the-art-details} provides details on these baselines. 
% In our experiments, the \textit{generic baseline} for this task is  CodeBERT~\citep{feng2020codebert} and GraphCodeBERT~\citep{guo2020graphcodebert}. Code2Seq~\citep{alon2019code2seq} and Code2Vec~\citep{alon2019code2vec} uses the same task to pre-train the model so we do not load the pretrained weights from the two models.
% }

\smallskip\noindent
\textit{Solution Classification:} 
% \revise{
This refers to the classification of source code into a predefined number of classes, e.g., based on the task it solves~\cite{pinter2018classification}, or programming languages~\cite{gilda2017source}. 
% The aim of this task is to classify which problem the code solve, which 
% Solution classification 
This is useful to assist or assess programming tasks and manage code warehouse. 
% For instance, to recognize the functionality of programs or automatically specify the task, algorithm or programming language(s) of published source code. 
% \todo{More details about solution classification, cite some papers here}. 
We evaluated all four generic baselines on this task, as well as three specialised learning-based approaches for this task, namely CNNSentence~\cite{ohashi2019convolutional}, 
OneCNNLayer~\cite{pinter2018classification}, SequentialCNN~\cite{gilda2017source} (\Cref{tab:state-of-the-art-details}).
% provides details about each baseline. 
We evaluated this task using the Java250 dataset,  which was designed %specifically 
for this task in previous studies~\cite{puri2021project} (\textit{see} \Cref{tab:subject_programs}). 

\smallskip\noindent
\revise{
\textit{Patch Classification:}
For this task, the aim %ain goal 
% of this task is 
is to identify the correctness of patches, i.e., if a patch is  
%the classification of 
(in)correct, wrong 
or over-fitting~\cite{ye2021automated,xiong2018identifying}. In our experiment, we compare the performance of \approach to the four generic baselines, as well as the current state-of-the-art learning-based approach for patch classification, i.e, ODS~\cite{ye2021automated}. %In this experiment, w
We employed the Defects4J~\cite{just2014defects4j} dataset (\textit{see} \Cref{tab:subject_programs}) which has also been used by previous studies for this task~\cite{ye2021automated,xiong2018identifying}. The goal of this task is to identify over-fitting APR patches. %In this experiment, w
We used five (5) programs and 890 APR patches\footnote{\revise{We exempted 12 patches out of the 902 patched programs used  by ODS, since they deleted complete functions, and there is no code representation for deleted functions.}} containing \revise{643} over-fitting patches and \revise{247} correct patches. 
}

\smallskip\noindent
\revise{
\textit{Mutant Prediction:} The goal of this task is to predict different types of mutants employed during mutation testing. Mutation testing is an important SE task that is typically deployed to determine the adequacy of a test suite to expose injected faults in a program~\citep{papadakis2019mutation}. In this work, we %setup two sub-tasks, namely (a) a \textit{binary classification task} to 
predict if a mutant is \textit{killable} or \textit{live}. 
% ,
% and (b) a \textit{multi-label classification task} to predict if a mutant triggers an \textit{exception}, exposes a \textit{fault}, leads to a \textit{timeout} or is \textit{live}. 
To this end, we employ the Defects4J~\cite{just2014defects4j} dataset (\textit{see} \Cref{tab:subject_programs}) which has been popularly employed for several SE tasks, including mutation testing~\citep{papadakis2019mutation}. We curated a mutant prediction dataset containing \revise{15} Java programs, and \revise{16,216} mutants. 
%We have chosen this dataset because it ....  
}

\smallskip\noindent
\textbf{Pre-training Setup:}  
% \revise{
For model pre-training, we curated the  \texttt{Jimp-\\le-Graph} dataset from the Maven repository\footnote{https://mvnrepository.com/},  it contains 1,976 Java libraries with about 3.5 millions methods in total. We randomly sample around 10\% data for the pre-traning purpose.
These Java libraries are from 42 application domains, this ensures a reasonable 
% to ensure the 
program diversity,
% in our %of the 
% pre-training dataset, 
these domains include math and image processing libraries. For the BiLSTM component (\Cref{sec:lstm}), we use one layer with hidden dimension size 150. We pre-train sub-tokens using the \texttt{Jimple} text for each program, the sub-token embedding dimension is set to 100 (\textit{see} \Cref{sec:approach}). We fine-tune the downstream tasks using the obtained pre-trained weights after one epoch. All GNNs use five (5) layers with dropout ratio 0.2. We use Adam~\citep{kingma2014adam} optimizer with 0.001 learning rate. In our experiment, we evaluated all three (3) pre-training strategies (\Cref{sec:pretraining}).
% }

% \todo{Metrics: add stat tests if we perform two-sample tests or odds ratio}

\smallskip\noindent
\textbf{Metrics and Measures:}  
%\todo{discuss F1 and accuracy and where each is used and why}
For all tasks, we report F1-score, precision and recall.
% \footnote{For instance, for method name prediction task (using Java-Small), we report the F1-score, precision and recall over the subtokens~\citep{10.1145/2786805.2786849, 10.1145/3296979.3192412} as the evaluation metrics. These three test metrics are widely used for this task~\citep{alon2019code2seq, alon2019code2vec, bui2021infercode}. They %se three metrics 
% are important to capture the correctness of a model's method name prediction, especially in relation to the original method name (aka ground truth). 
% % \revise{
% For example, consider a function with original method named \textit{registerUser}, if a model predicts its name as 
% % a method name model prediction 
% \textit{userRegister}, then this is considered an exact match. However, 
% % method name prediction 
% if a model's prediction is \textit{user}, it has full precision but low recall. Meanwhile, if the model's prediction is \textit{registerAllUser}, then it has full recall but low precision.}
We discuss most of our results using F1-score since it is the harmonic mean of precision and recall. Besides, 
% it is ,  this, considering 
it is a better measurement metric than accuracy, especially when the dataset is imbalanced (e.g., Java-Small). Hence, we do not report the accuracy for imbalanced datasets, e.g., mutant data  is imbalanced % dataset 
with about $30\%$ live mutants and $70\%$ killable mutants. 
% In addition, 
%However, for solution classification, 
% we also report F1-score, precision, recall. 
% and Accuracy because the employed dataset (Java250) is balanced. 
% Besdies, 
%we report the model accuracy for this task since the employed dataset (Java250) is balanced, i.e.,  %is used when 
%the class distribution is similar.
%  % it considers true positive and true negative. 
% For Code Clone Detection on GCJ, we report the F1-score, Precision and Recall, similar to other clone detection works~\citep{deepsim, guo2020graphcodebert, wang2020detecting}. We do not report the accuracy because the employed datadet (GCJ) is imbalanced, % dataset 
% with about $20\%$ similar code pair and $80\%$ non-similar code pair. 
%  
%We provide more evaluation details and results in the supplementary material and Github repository\footnote{\url{https://github.com/graphcode2vec/graphcode2vec}}.
We provide the code details in the Github repository\footnote{\url{https://github.com/graphcode2vec/graphcode2vec}}.
% }
%\smallskip\noindent
%\textit{Accuracy:}\todo{XXX}
%\smallskip\noindent
%\textit{F1-Score:}\todo{XXX}

\begin{table*}[bt]
  \caption{ %\centering 
  Effectiveness of \approach vs. 
  Syntax-only Generic Code Embedding approaches. The best results 
  %for %each task/metric 
  are in \textbf{bold} text, the results for the best-performing baseline 
%   (syntax-only approach) 
  are in \textit{italics}. We report the improvement in effectiveness between \approach and the best-performing baseline in ``\% Improvement'', improvements above five percent (>5\%) are in \textbf{bold} text.
%   \todo{IS f1 vs. acc an issue here? esp. since where we report F1 are the only places we perform better, think it is better/safe to report both or one metric. I think for F1, using accuracy is better than using recall and precision. I will evaluate them again using the saved best model.}
}
\vspace{-0.35cm}
 \scalebox{0.95}{
\begin{tabular}{|l|ccc|ccc|ccc|%ccc|
ccc|} %c|c|} %l|} 
\hline
% \multirow{2}{*}{
\textbf{Generic Code}
% }  
& \multicolumn{3}{c|}{\textbf{Method Name Prediction}} & \multicolumn{3}{c|}{\textbf{Solution Classification}} & 
% \multicolumn{3}{c|}{\textbf{Code Clone Detection}} & 
\multicolumn{3}{c|}{\textbf{ %Multi-label 
Mutant Prediction % / (Binary)
}} &  \multicolumn{3}{c|}{\textbf{Patch Classification}}
\\ 
\textbf{Embedding} & \textbf{F1} & \textbf{Preci} & \textbf{Recall} &  \textbf{F1} & \textbf{Preci} & \textbf{Recall} & \textbf{F1} & \textbf{Preci} & \textbf{Recall} & \textbf{F1} & \textbf{Preci} & \textbf{Recall}  
% & \textbf{F1} & \textbf{Preci} & \textbf{Recall} 
\\ \hline
Code2Seq & \textit{0.4920} & \textit{0.5963} & 0.4187 & 0.7542 & 0.7678 & 0.7536 & 
% 0.9883 & 0.9886 & 0.9882 & 
% 0.3683 / (
0.5911 & %0.3728 / (
0.6423 & %0.3694 / (
0.5881 & 0.8901 &  0.8355 & \textit{0.9541} 
\\ 
Code2Vec & 0.3309 & 0.3779 & 0.2943 & 0.8034 & 0.8081 & 0.8028 & 
% 0.9947 & 0.9940 & 0.9948 & 
% 0.3780 / (
0.6398 &  %0.3819 / (
0.6632 & %0.3787 / (
0.6320  &  0.8787 & 0.8806 & 0.8782 
\\
CodeBERT  & 0.3963 & 0.3295 & \textit{0.4969} & \textit{0.8783} & \textit{0.8747} & \textit{0.8878} & 
% \textbf{0.9980} & \textbf{0.9986} & \textbf{0.9973} & 
% \textit{0.4491} / (
\textit{0.7106} & %\textit{0.4508} / (
\textit{0.7305} & %\textit{0.4508} / (
\textit{0.6995} & \textit{0.9275} &  \textit{0.9099} & 0.9473 
\\
% \hline 
% GraphCodeBERT & 0.5761 & \textbf{0.7261} & 0.4775 & \textbf{0.9850} & \textbf{0.9868} & \textbf{0.9843} & \textbf{0.9990} & \textbf{0.9991} & \textbf{0.9989} &  \textbf{0.7680} & \textbf{0.7623} & \textbf{0.7649} &  0.9108 & 0.9557 & 0.9317
%  \\
\hline
% \approach (GAT+Node)   
\approach %(GAT+Context) 
& \textbf{0.5807} & \textbf{0.6150} & \textbf{0.5502} & \textbf{0.9746} & \textbf{0.9753} & \textbf{0.9746} & 
% \textit{0.9971} & \textit{0.9971} & \textit{0.9971} & 
% \textbf{0.5021} / (
\textbf{0.7542} &  %\textbf{0.5019} / (
\textbf{0.7569} & %\textbf{0.5026} / (
\textbf{0.7524}  & \textbf{0.9359}
&  \textbf{0.9145} & \textbf{0.9602} 
\\
\hline
 \% Improvement & \textbf{18.03\%} & 3.14\%  & \textbf{10.73\%} & \textbf{10.96\%} & \textbf{11.50\%} & \textbf{9.78\%} & 
%  -0.09\% & -0.15\% & -0.02\% & 
%  \textbf{11.80\%} / (
 \textbf{6.14\%} & %\textbf{11.34\%} / (
 3.61\% & %\textbf{11.91\%} / (
 \textbf{7.56\%}   & 0.91\% & 0.51\% & 0.64\%
 \\
% & \textbf{0.6194}  & 0.9752  & \textbf{0.9993} \\ 
\hline
% \textbf{Average} & & & & \\ \hline
% \textbf{Average} & & & & \\ \hline
\end{tabular}
}
\vspace{-0.35cm}
\label{tab:results-generic-code-embedding}
\end{table*}

\smallskip\noindent
 \textbf{Probing Analysis:} % and Ablation Study}  
%  \revise{
The goal of our probing analysis is to ensure that lexical and dependence features are indeed learned by \approach's code embedding. Probing is a widely used technique to examine an embedding for desired properties~\cite{bertology, conneau-etal-2018-cram, pse}. To this end, we trained diagnostic classifiers to probe \approach's code embedding for our desired properties (i.e., lexical and/or program dependence features). Concretely, we train a simple classifier with one MLP layer fed with 
the learned code embedding (e.g. lexical) to examine if our code embedding encodes the desired property. To achieve this, we curated a dedicated dataset for training and evaluating our probing classifiers. Specifically, we employ three probing datasets, namely LeetCode-10, M-LeetCode and Concurrency (\Cref{tab:subject_programs}). We have employed these datasets because they require lexical or dependence embedding to address their corresponding tasks. 

\smallskip\noindent
\textbf{Probing Task Design:}
\label{sec:probingexperiments}
% \revise{
We design four probing tasks.
The first three (Task-1, Task-2 and Task-3) use 
 LeetCode-10 and M-LeetCode, 
and the last one (Task-4) 
uses Concurrency.
\emph{Task-1} classifies what problem the
solution code solves on LeetCode-10. LeetCode-10 shares lexical token similarities within one problem group, and some solutions from the different problem groups may have the same semantic structure, e.g., using one for-loop. Therefore,  we hypothesize that the lexical embedding  is more informative than the semantic embedding for Task-1. 
\emph{Task-2} mixes LeetCode-10 and M-LeetCode,
and then judges which dataset the input code is from (binary
classification). LeetCode-10 and M-LeetCode share lots of similar lexical tokens but the code semantic structures are different. Hence, the semantic embedding should be more informative  than the code lexical syntactic embedding. 
\emph{Task-3} also mixes the two datasets but uses all the 20 labels
instead of a binary classification. Task-3 integrates Task-1 and Task-2, requiring both lexical and semantic information.
\emph{Task-4} is a concurrency bug classification task. The code with same label can have the high lexical similarity but the code semantic structure should be different.
% }

% \revise{
% \smallskip\noindent
%  \textbf{Ablation Study Design:}
% %Probing analysis shows the different characteristics of the program
% %token feature and the program structure feature.
% To evaluate the
% practical impacts of the lexical embedding and dependencce
% embedding on downstream tasks, we fine-tune the model
% for method name prediction (Java-Small) and solution classification (Java250) tasks, using %with the 
% two types of program-encoded feature individually. 
% % }
% % revise{
% The goal of this study is to investigate 
% % We examine how removing lexical embedding or dependencce em-bedding during the fine-tuning ofGraphCode2Vec’s pre-trainedmodel 
% % Ablation analysis can 
% % reveal 
% % how the pre-training strategies influence
% % the expressiveness of GNN, and give the evidence that 
% how the lexical embedding 
% % representation 
% and dependencce embedding of \approach affect its effectiveness 
% % of \approach 
% on downstream tasks. 
% are complementary.
% It also gives some
% tips how  we should build up deep learning models for Software
% Engineering tasks.
% }

\smallskip\noindent
\textbf{\approach's Configuration:}
% \revise{
% In our evaluation, w
We employ three (3) pre-training strategies, namely node classification, context prediction and VGAE.
% (\Cref{sec:pretraining}). 
Our approach supports 
% allows to employ one of  
% We uses 
four (4) GNN architectures for dependence embedding (\textit{see} \Cref{sec:approach}), namely GCN~\citep{kipf2016semi}, GraphSAGE~\citep{hamilton2017inductive}, GAN~\citep{velivckovic2017graph} and GIN~\citep{xu2018powerful}. In total, we have 12 %($3 \times 4$) 
possible configurations. However, the % and the 
\textit{default configuration is context prediction for pre-training and dependence embedding with GAT architecture}. 
% \Cref{sec:approach} describes each GNN configuration and pre-training strategy. 
In our experiments, we evaluate the effect of 
% contribution of 
each configuration on the effectiveness of our approach (\textit{see} \Cref{sec:results}). 
% Context-Prediction is using the neighbors to predict the center node representation while Instruction Classification is to predict the node label and VGAE is to judge if the two nodes are linked or not. Intuitively, Context-Prediction is more general task than the other tow pretraining methods. GAT uses attention mechanism that can lead to the embedding with more weights about the important information.
% }

% \smallskip\noindent
% \textbf{Research Protocol:}  
% \todo{XXXX}

\smallskip\noindent
\textbf{Implementation Details and Platform:}
% \revise{ 
\approach was implemented in about 4.8~KLOC of Python code, using the Pytorch ML framework. 
Our data processing and evaluation code is about 
% We also implement 
3~KLOC of Java code. 
% for data processing. 
We use Soot \citep{vallee2010soot} to extract the program dependence graph (PDG). We reuse the code from the public repository of each baseline in our experiments.\footnote{https://github.com/tech-srl/code2vec,https://github.com/tech-srl/code2seq, https://github.com/microsoft/CodeBERT, https://github.com/hukuda222/code2seq} However, we adapt each baseline to our downstream tasks, e.g., by replacing the classifier but using the same performance metrics. 
%For all approaches, we use one multilayer perceptron layer as the decision layer. 
All experiments were conducted on a Tesla V100 GPU server, with 40 CPUs (2.20 GHz) and 256G of main memory. The implementation of \approach is available online\footnote{https://github.com/graphcode2vec/graphcode2vec}.

% \begin{center}
% \url{}    
% \end{center}

% }

% \textbf{based approaches} & \textbf{Accuracy}  & \textbf{F1-Score} & \textbf{Recall}  & \textbf{Precision}  \\ \hline % & \textbf{Average} \\ 
% ODS & 0.8865 & 0.8995 & 0.9502 & 0.9238
%  \\
% \hline
% \approach &\textbf{0.9045}
%  & \textbf{0.9145} & \textbf{0.9602} & \textbf{0.9359} \\

%\input{results}
\section{Experimental Results}
\label{sec:results}

\begin{table}[bt]
  \caption{ %\centering 
  Effectiveness of \approach (aka ``\textsc{Graph.}'') vs. Task-Specific learning-based approaches for two SE tasks. The best results are in \textbf{bold} text, the results for the second best-performing approach are in \textit{italics}. The improvement in effectiveness between \approach and the best-performing baseline is reported in %column 
  ``\textsc{Graph.} \textit{(\% Improv.)}''.
}
\vspace{-0.35cm}
 \scalebox{0.6}{
\begin{tabular}{|l|%ccccc|c|
ccc|c|cccc|c|} 
\hline
% \multirow{2}{*}{
% \textbf{Generic Code}
% }  

& %\multicolumn{6}{c|}{\textbf{Code Clone Detection}} & 
\multicolumn{4}{c|}{\textbf{Solution Classification}} &   \multicolumn{5}{c|}{\textbf{Patch Classification}} %& \multicolumn{1}{c|}{\textbf{\% Improv.}}
\\ %\hline
% & & \textsc{Graph}
%  & \textbf{DECK-} 
%  \multirow[c]{1}{*}{\rotatebox[origin=c]{90}{\parbox{0.1cm}{\textbf{DECK-}}}}
%  & \textbf{RTV-} & \textbf{Deep-} & \textbf{FA-} & \textbf{FA-} & \multirow{1}{*}{\textbf{\textsc{Graph.}}} & 
 & \textbf{CNN} & \textbf{One} & \textbf{Seq.-} & \multirow{1}{*}{\textbf{\textsc{Graph.}}} & \textbf{SimFea-} & \textbf{Prop-} & \textbf{Patch-} & \multirow{2}{*}{\textbf{ODS}} & \multirow{1}{*}{\textbf{\textsc{Graph.}}}
\\
% & 
% \textbf{ARD} 
% & \textbf{NN} & \textbf{Sim} & \textbf{AST$^1$} & \textbf{AST$^2$} & \textit{(\% Improv.)} %\textbf{\textsc{Code2Vec}} 
&  \textbf{Sen.} & \textbf{CNN.} & \textbf{CNN} & \textit{(\% Improv.)} %\textbf{\textsc{Code2Vec}} 
& \textbf{tures} & \textbf{het} &  \textbf{Sim} & & \textit{(\% Improv.)}  %\textbf{\textsc{Code2Vec}}
\\ \hline
 \textbf{F1-Score} 
%  & 0.450 & 0.200 & 0.710 & 0.960 & \textit{0.990} &  \textbf{0.997 \revise{(7.1\%)}} 
 & \textit{0.690} & 0.540 & 0.470 & \textbf{0.970 \revise{(40.6\%)}} & 0.881 & 0.892 & 0.881 & \textit{0.900} & \textbf{0.915 \revise{(1.7\%)}}
 \\ \hline
 \textbf{Recall} 
%  & 0.440 & 0.900 & 0.820 & \textbf{1.00} & 0.970 & \textit{0.997 \revise{(-0.3\%)}} 
 & \textit{0.690} & 0.540 & 0.470 & \textbf{0.970 \revise{(40.6\%)}} & 0.895 & 0.891 & 0.389 & \textit{0.950} & \textbf{0.960 \revise{(2.1\%)}}
 \\ \hline
\textbf{Precision} 
% & 0.440 & 0.330 & 0.760 & 0.970 & \textit{0.980} &  \textbf{0.997 \revise{(1.7\%)}} 
& \textit{0.700} & 0.550 & 0.480 & \textbf{0.970 \revise{(38.6\%)}} & 0.870 &  0.889 & 0.830 & \textit{0.924} & \textbf{0.936 \revise{(1.3\%)}}
 \\ \hline
\end{tabular}
}
%\vspace{-1cm}
\label{tab:results-task-specific-approaches}
\end{table} 

\smallskip\noindent
\textbf{RQ1 Task-specific learning-based applications:} 
% \revise{  
This experiment examines how \approach compares to \recheck{seven (7)} state-of-the-art task-specific learning-based %applications
% . % for code clone detection task. 
% We compare the performance of \approach to that of %the state-of-the-art task-specific 
% learning-based approaches that are specialised 
techniques for %\textit{code clone detection}, 
\textit{solution classification} and \revise{\textit{patch classification}. We selected these two tasks %solution classification and code clone detection 
for this experiment 
due to their popularity, availability of ML-based baselines and their application to vital %many other 
% software engineering
SE
tasks, e.g., automated program repair, patch validation, code evolution, %program search 
and software warehousing. }%, refactoring and reuse.}
% For solution classification, we use  5 code embedding methods from ProjectCodeNet~\citep{puri2021project}, i.e., CNN with token
% sequence and four types of GNNs(GIN, GIN-V, GCN, and GCN-V) with Simplified Parse Tree (SPT)~\citep{parr2013definitive}.
% For code clone detection, 
% In this experiment, we employ all four %task-specific 
% baselines for code clone detection %highlighted 
% in \Cref{tab:state-of-the-art-details}, namely DECKARD, FA-AST, DeepSim and RTvNN. 
% We compared to the two GNN configurations of FA-AST, namely Gated Graph Neural Network (GGNN)~\citep{li2015gated} and Graph Matching Network (GMN)~\citep{li2019graph}, denoted by FA-AST$^1$ and FA-AST$^2$, respectively.
% In addition, we compared \approach to one generic approach that has been trained and evaluated on clone-detection, i.e., ASTNN, making it six baselines in total.
% Overall, we compared \approach to four (4) baselines for this task. 
% Meanwhile, f
%cation, 
We evaluated against three solution classifiers, %task-specific baselines for solution classification 
namely CNNSentence~\cite{ohashi2019convolutional}, OneCNNLayer~\cite{pinter2018classification}, SequentialCNN~\cite{gilda2017source}. 
\revise{We also compare \approach to four %state-of-the-art 
patch classifiers -- %, namely 
Prophet~\cite{long2016automatic}, PatchSim~\cite{xiong2018identifying}, SimFeatures~\cite{wang2020automated} and ODS~\cite{ye2021automated}.% as the task-specific baseline for patch classification.
} 
%For this experiment, we employ the default configuration for our approach (\approach), \textit{i.e.}, \approach's GAT graph neural network with context pre-training. 
% }

\begin{table*}[bt]
  \caption{ %\centering 
  Effectiveness of \approach vs. 
  GraphCodeBERT. Lower complexity, the best results and higher improvements (above five percent (>5\%)) are in \textbf{bold} text.)
  %for %each task/metric 
  are in \textbf{bold} text. 
%   , the results for the best-performing baseline 
% %   (syntax-only approach) 
%   are in \textit{italics}.
% %   \todo{IS f1 vs. acc an issue here? esp. since where we report F1 are the only places we perform better, think it is better/safe to report both or one metric. I think for F1, using accuracy is better than using recall and precision. I will evaluate them again using the saved best model.}
}
\vspace{-0.35cm}
 \scalebox{0.85}{
\begin{tabular}{|l|cc|ccc|%ccc|
ccc|ccc|ccc|} %c|c|} %l|} 
\hline
% \multirow{2}{*}{
\textbf{Generic Code} 
&
\textbf{Model} & \textbf{Pretrain}
% }  
&  \multicolumn{3}{c|}{\textbf{Method Name Prediction}} & \multicolumn{3}{c|}{\textbf{Solution Classification}} & 
%\multicolumn{3}{c|}{\textbf{Code Clone Detection}} & 
\multicolumn{3}{c|}{\textbf{Mutant Prediction}} &  \multicolumn{3}{c|}{\textbf{Patch Classification}}
\\ 
\textbf{Embedding} & \textbf{Size} & \textbf{Data} & \textbf{F1} & \textbf{Preci} & \textbf{Recall} &  \textbf{F1} & \textbf{Preci} & \textbf{Recall} & \textbf{F1} & \textbf{Preci} & \textbf{Recall} & \textbf{F1} & \textbf{Preci} & \textbf{Recall}  
% & \textbf{F1} & \textbf{Preci} & \textbf{Recall} 
\\ \hline

% \hline 
GraphCodeBERT & 124M & 2.3M & 0.5761 & \textbf{0.7261} & 0.4775 & \textbf{0.9850} & \textbf{0.9868} & \textbf{0.9843} & 
% \textbf{0.9990} & \textbf{0.9991} & \textbf{0.9989} & 
% \textbf{0.5345}/(
\textbf{0.7649} & %\textbf{0.5364}/(
\textbf{0.768} & %\textbf{0.5347}/(
\textbf{0.7623} & 0.9317 &  0.9108 & 0.9557 
  \\
\hline
% \approach (GAT+Node)   
\approach %(GAT+Context) 
& \textbf{2.8M} & \textbf{314K} & \textbf{0.5807} & 0.6150 & \textbf{0.5502} & 0.9746 & 0.9753 & 0.9746 & 
% \textit{0.9971} & \textit{0.9971} & \textit{0.9971} & 
% \textit{0.5021}/(
\textit{0.7542} & 
% \textit{0.5019}/(
\textit{0.7569} & 
% \textit{0.5026}/(
\textit{0.7524}  & \textbf{0.9359}
&  \textbf{0.9145} & \textbf{0.9602} 
\\
\hline
 \% Improvement & \textbf{50X} & \textbf{7X}  & \textbf{7.99\%} & -15.30\% & \textbf{15.23\%} & -1.07\% & -1.17\% & 
%  -0.99\% & -0.19\% & -0.20\% & 
 -0.18\% & 
%  -6.06\%/(
 -1.40\% & %-6.43\%/(
 -1.45\% & %-6.00\%/(
 -1.30\% & 0.45\% & 0.41\% &0.47\%   
\\
\hline
\end{tabular}
}
\vspace{-0.35cm}
\label{tab:results-generic-code-embedding-graphbert}
\end{table*}
% \todo{update discussion and results with that of patch classification}

\revise{
Our evaluation results show that \textit{\approach outperforms the state-of-the-art task-specific learning based approaches for the tested tasks, i.e., patch classification, %code clone detection 
and solution classification}. 
% \Cref{tab:results-task-specific-embedding}, \Cref{tab:results-task-specific-embedding-patch} and  \Cref{tab:results-task-specific-embedding-solution-classification} 
\Cref{tab:results-task-specific-approaches} highlights the effectiveness of \approach in comparison to learning-based approaches for %code clone detection, 
patch classification and solution classification, respectively. In particular, \approach outperforms 
all seven task-specific baselines in our evaluation. 
% \revise{
% Specifically, 
% our approach is between two to three times more effective than  DECKARD and RtvNN (\textit{see} \Cref{tab:results-task-specific-approaches}). 
% In addition, it slightly outperforms FA-AST across all tested measures. Besides our approach, FA-AST is the best performing learning-based 
% baseline in our code clone detection evaluation. \approach is up to seven percent more effective than the most effective clone detection baseline -- FA-AST$^2$ (\textit{see} \Cref{tab:results-task-specific-approaches}). 
% Likewise, 
\approach outperforms all three baselines for solution classification, it is almost twice as effective as SequentialCNN and OneCNNLayer, and 40\% more effective than the best baseline -- CNNSentence (\textit{see} \Cref{tab:results-task-specific-approaches}). 
% }
% \revise{
In addition, \approach outperforms all four state of the art %learning-based 
patch classifiers, %cation methods, 
i.e., ODS~\cite{ye2021automated}, Prophet~\cite{long2016automatic}), PatchSim~\cite{xiong2018identifying} and SimFeatures~\cite{wang2020automated}. 
% \approach is more than 
It is at least twice %up to two times 
as effective as PatchSim (in terms of recall) and slightly (\recheck{up to 2}\%) more effective than the best baseline, i.e., ODS (\textit{see} \Cref{tab:results-task-specific-approaches}). 
% }
This result demonstrates the utility of our approach in addressing %(these two) 
both downstream tasks. 
% results
Furthermore, %these results 
it highlights the effectiveness of generic code embedding in comparison to specialised learning-based approaches.
% , %this is 
% evident by the 
% % model pre-training. 
% % further shows that \approach outperforms specialised learning methods for 
% effectiveness of \approach on patch classification, code clone detection and solution classification. %We attribute this superior performance to 
This superior performance can be attributed to the fact that \approach is generic, %captures semantic information (program dependence) 
and it employs self-supervised %benefits from %the influence of 
model pre-training. 
}

\begin{result}
% \revise{
% In our evaluation, 
\approach is up to two times (2x) more effective than %is comparable to 
% outperforms 
the \revise{seven (7)} state-of-the-art 
task-specific %specialised 
% learning-based tested
approaches, for both tasks. 
% /by up to three times. % for code clone detection.
% it outperforms \todo{X out of five} baselines. 
% for code clone detection.  
% }
\end{result}

\smallskip\noindent
\textbf{RQ2 Generic Code embedding:} 
% \todo{Add text to describe Accuracy of Solution Classification.}
% \revise{
% The goal of 
In this experiment, we demonstrate how \approach compares to the state-of-the-art generic code embedding approaches. 
% In this experiment, we examine the \textit{effectiveness} of our approach (\approach), in comparison to the state-of-the-art generic code embedding approaches. 
\revise{%In particular,0 
We thus, compare the effectiveness of \approach with three (3) syntax-only generic baselines, namely CodeBERT, Code2Seq and Code2Vec. 
Additionally, we compare the effectiveness of our approach to a a larger and more complex state-of-the-art generic approach that captures both syntax and semantics, specifically, GraphCodeBERT. %, %, combined syntactic and semantic -based approach, namely . 
%These baselines represent the major of generic code embedding. 
% , specifically, syntax-only approaches (\textit{i.e.}, CodeBERT, Code2Seq and Code2Vec) as well as approaches that capture both syntax and semantics (\textit{\textit{i.e.}}, GraphCodeBERT). }
% In this experiment, w
We used %experimented on %, using 
% \revise{
four (4) downstream SE tasks -- %namely 
method name prediction, solution classification, 
% code clone detection, 
mutant prediction and patch classification.} 
% In our evaluation of method name prediction task, we employ the 
% % For the evaluation of 
% Code2Vec and Code2Seq models provided by the original papers. 
% % for the method name prediction task, we employ the evaluated trained model for the task. 
% For all other approaches and tasks, we employed their pre-trained models, then fine-tuned for the downstream task at hand. We used the default GNN and pre-training configurations for our approach,
% % of our approach, 
% \textit{i.e.}, \approach's graph neural network GAT with context pre-training. 
% }

\noindent
\revise{
\textit{\textbf{Syntax-only Generic Embedding}:}
In our evaluation, we found that \textit{our approach (\approach) outperforms all syntax-based generic baselines for all tasks}. \Cref{tab:results-generic-code-embedding} highlights the effectiveness of \approach in comparison to the baselines (i.e., Code2seq, Code2Vec and CodeBERT). As an example, consider method name prediction, \approach is twice 
as effective as some baselines, e.g., Code2Vec. 
For \recheck{all (four) tasks}, %(four out of five)}, 
% \approach is more effective than the best baseline, it is up to 18\% more effective than the best baseline in method name prediction. 
% For all tasks, %(except code clone detection), 
\approach clearly outperforms all baselines across all metrics. It is up to 12\% and 18\% more effective than the best baselines, CodeBERT and Code2Seq, respectively. We observed CodeBERT is the best baseline %, 
% outperforming \approach in one task -- code clone detection. CodeBERT is also 
%it is the second best approach 
on three tasks. 
% , and 
% it slightly outperforms \approach in code clone detection. %\footnote{
% and comparable to . 
We attribute the performance of CodeBERT on these tasks to its much higher complexity (i.e., huge number of trainable parameters, more than \recheck{124M}) and the size of the pre-training dataset (\recheck{8.5M})~\cite{husain2019codesearchnet}. %} 
Overall, our results demonstrate that including semantic program features improves the performance of code representation across these downstream tasks. Thus, emphasizing the importance of semantic features in addressing SE tasks, especially the need to %precisely 
capture program dependencies %information 
in code representation.
}

% \Cref{tab:results-generic-code-embedding} highlights the effectiveness of \approach in comparison to the baselines. 
% Our approach is up to \todo{X} time more effective than the best performing syntax-based generic baseline (i.e., \todo{X}). 

\begin{result}
\revise{
For all (four) tasks, \approach is (up to 18\%) more effective than (the best) syntax-only baselines. %, by up to 18\%.  
% more effective than the best %syntax-based generic 
% baselines. 
}
\end{result}

\noindent
\revise{
\textit{\textbf{Complementarity with GraphCodeBERT:}}
% \textbf{
% Combined Syntactic and Semantic -based Generic Embedding}:
We also observe that despite the lower complexity of our approach (\approach), \textit{it is comparable and complementary to GraphCodeBERT} across tested tasks. 
% GraphCodeBERT is 
% a more complex approach than \approach. S
GraphCodeBERT captures both syntactic and semantic program features but, it is significantly larger and complex than \approach. 
\recheck{\Cref{tab:results-generic-code-embedding-graphbert}} highlights the complexity and effectiveness of GraphCodeBERT in comparison to \approach. 
For instance, GraphCodeBERT has at least 50 times (50x) as many trainable parameters as \approach (124 million versus 2.8 million parameters), and seven times (7x) as much pre-training data (2.3M versus 314K methods). 
Despite the difference in size and complexity, GraphCodeBERT has a comparable performance to \approach. Specifically, \approach outperforms GraphCodeBERT on two tasks (\recheck{method name prediction and patch classification}) and it is comparable on the other \recheck{two} tasks (\recheck{solution classification, %code clone detection 
and mutant prediction}). Notably, GraphCodeBERT has a negligible improvement over \approach for these two tasks (about 1\%). 
These results demonstrate that although simpler and trained on 7 times less  data, \approach is complementary to GraphCodeBERT. This disparity in size and complexity implies that precise program dependence information is important. Nevertheless, our results  show that both \approach and GraphCodeBERT are more effective than syntax-only approaches, e.g., CodeBERT (\textit{cf.}  \Cref{tab:results-task-specific-approaches} and \Cref{tab:results-generic-code-embedding-graphbert}). 
}

\begin{result}
\revise{
\approach is complementary to GraphCodeBERT despite being simpler and trained on seven times (7x) less data. It is more effective on two tasks, and comparable on the other \recheck{two} tasks. 
}
\end{result}

% \item 

% \begin{table}[!bt]
%   \caption{Solution Classification: Effectiveness of \approach vs. %the state-of-the-art 
% %   task-specific 
%   specialised 
% %   learning methods 
% %   -based applications 
% ML methods. Best results are \textbf{bold}.}
% %   \vspace{-0.3cm}
% \vspace{-1em}
%  \scalebox{0.8}{
% \begin{tabular}{|l|l|l|l|l|l|} %l|} 
% \hline
% \textbf{Task-specific learning-} & \multicolumn{4}{c|}{\textbf{Solution Classification}} \\ % & \textbf{Approach} \\  
% \textbf{based approaches}&\textbf{Accuracy} & \textbf{F1-Score} & \textbf{Recall}  & \textbf{Precision}  \\ \hline % & \textbf{Average} \\ 
% CNNSentence & 0.69 & 0.69 & 0.69 & 0.70 \\
% OneCNNLayer & 0.54 & 0.54 & 0.54 & 0.55 \\
% SequentialCNN & 0.47  & 0.47 & 0.47 & 0.48 \\
% \hline
% \approach & \textbf{0.98} & \textbf{0.97} & \textbf{0.97} & \textbf{0.97} \\
% % &  0.99 \\ 
% \hline
% \end{tabular}
% }
% % \vspace{-1.17cm}
% \vspace{-2em}
% \label{tab:results-task-specific-embedding-solution-classification}
% \end{table}

\begin{table*}[!t]
\centering
\caption{Probing Analysis results showing the accuracy for all pre-training strategies and GNN configurations. 
% , using frozen pre-trained models for  embedding and semantic (program dependence) embedding. 
Best results for each sub-category are in bold, and the better results between  syntactic (lexical) embedding and semantic embedding is in \textit{italics}. ``syn+sem'' refers to \approach's models capturing both syntactic and semantic features.}
\vspace{-1em}
\label{tab:probingTask}
\scalebox{0.8}{
\begin{tabular}{|l|l|cccc|cccc|cccc|cccc|}
\hline
% \multicolumn{2}{|l|}{\multirow{2}{*}{}} 
\textbf{Pre-training} & \textbf{Captured} & \multicolumn{4}{c|}{\textbf{Task-1} (syntax-only) }           & \multicolumn{4}{c|}{\textbf{Task-2} (semantic-only)}   & \multicolumn{4}{c|}{\textbf{Task-3} (syntax and semantic)}   & \multicolumn{4}{c|}{\textbf{Task-4} (semantic-only)}   \\ %\cline{3-18} 
% \multicolumn{2}{|l|}{}                  
\textbf{Strategy} & \textbf{Feature} & \textbf{GCN}     & \textbf{GIN}     & \textbf{GSAGE}   & \textbf{GAT}     & \textbf{GCN}   & \textbf{GIN}   & \textbf{GSAGE} & \textbf{GAT}   & \textbf{GCN}   & \textbf{GIN}   & \textbf{GSAGE} & \textbf{GAT}   & \textbf{GCN}   & \textbf{GIN}   & \textbf{GSAGE} & \textbf{GAT}   \\ \hline
\multirow{3}{*}{\textbf{Context}}    & \textbf{semantic}    & $0.822$ & $0.674$ & $0.842$ & $0.886$ & \textbf{\textit{0.684}} & \textit{0.614} & \textbf{\textit{0.704}} & \textit{0.741} & $0.513$ & $0.381$ & \textit{0.543} & \textbf{\textit{0.612}} & \textbf{\textit{0.654}} & \textbf{\textit{0.666}} & \textbf{\textit{0.657}} & \textbf{\textit{0.594}} \\ 
                            & \textbf{syntactic}     & \textbf{\textit{0.934}} & \textbf{\textit{0.938}} & \textit{0.942} & \textit{0.928} & $0.615$ & $0.602$ & $0.617$ & $0.602$ & \textit{0.529} & \textit{0.527} & \textit{0.528} & $0.527$ & $0.580$ & $0.525$ & $0.524$ & $0.449$ \\ %\hline
                            & \textbf{syn+sem} & $0.918$ & $0.928$ & \textbf{0.95}  & \textbf{0.942} & $0.641$ & \textbf{0.641} & $0.688$ & \textbf{0.797} & \textbf{0.559} & \textbf{0.546} & \textbf{0.587} & $0.6$   & $0.605$ & $0.592$ & $0.608$ & $0.592$ \\ \hline
\multirow{3}{*}{\textbf{Node}}       & \textbf{semantic}    & $0.758$ & $0.820$ & $0.802$ & $0.840$ & \textbf{\textit{0.651}} & \textbf{\textit{0.667}} & \textbf{\textit{0.741}} & \textbf{\textit{0.686}} & $0.426$ & \textbf{\textit{0.514}} & \textit{\textbf{0.625}} & \textit{\textbf{0.563}} & \textbf{\textit{0.647}} & \textbf{\textit{0.664}} & \textbf{\textit{0.659}} & \textbf{\textit{0.670}} \\ 
                            & \textbf{syntactic}     & \textbf{\textit{0.904}} & \textit{0.884} & \textbf{\textit{0.876}} & \textbf{\textit{0.916}} & $0.584$ & $0.587$ & $0.606$ & $0.593$ & \textit{0.516} & $0.504$ & $0.490$ & $0.513$ & $0.484$ & $0.476$ & $0.420$ & $0.550$ \\ 
                            & \textbf{syn+sem} & $0.872$ & \textbf{0.9}   & \textbf{0.876} & $0.902$ & $0.624$ & $0.618$ & $0.691$ & $0.67$  & \textbf{0.522} & $0.508$ & $0.572$ & $0.545$ & $0.519$ & $0.522$ & $0.451$ & $0.57$  \\
                            \hline
\multirow{3}{*}{\textbf{VGAE}}       & \textbf{semantic}    & $0.856$ & $0.812$ & $0.868$ & $0.866$ & \textbf{\textit{0.594}} & \textbf{\textit{0.653}} & $0.583$ & \textbf{\textit{0.617}} & $0.403$ & \textit{0.532} & $0.407$ & $0.477$ & \textbf{\textit{0.673}} & \textbf{\textit{0.680}} & \textbf{\textit{0.674}} & \textbf{\textit{0.656}} \\ 
                            & \textbf{syntactic}     & \textit{0.916} & \textbf{\textit{0.932}} & \textbf{\textit{0.928}} & \textbf{\textit{0.950}} & $0.591$ & $0.572$ & \textbf{\textit{0.594}} & $0.599$ & \textit{0.485} & $0.494$ & \textit{0.492} & \textbf{\textit{0.495}} & $0.523$ & $0.617$ & $0.584$ & $0.591$ \\ 
                            & \textbf{syn+sem}  & \textbf{0.92}  & $0.926$ & \textbf{0.928} & $0.938$ & $0.59$  & $0.63$  & $0.591$ & $0.596$ & \textbf{0.498} & \textbf{0.548} & \textbf{0.508} & $0.492$ & $0.627$ & $0.658$ & $0.531$ & $0.586$ \\ \hline
 \multicolumn{2}{|c|}{\textbf{Best Config.}} & \multicolumn{4}{c|}{\textbf{Syntactic = 8/12}} & \multicolumn{4}{c|}{\textbf{Semantic = 9/12}} & \multicolumn{4}{c|}{\textbf{Syntactic + Semantic =  7/12}} & \multicolumn{4}{c|}{\textbf{Semantic = 12/12}} \\ \hline
\end{tabular}}
\vspace{-1em}
\end{table*}

% \begin{table}[!ht]
% \centering
% \caption {Effectiveness of \approach versus the state-of-the-art Task-specific learning-based approaches for Java250 Solution Classification}
% \vspace{-1em}
% \tablabel{tab:baseline_solution_classification}
% \scalebox{0.85}{
% \begin{tabular}{|l|l|l|l|l|l|l|}
% \hline
%       & CNN   & GCN & GCN-V & GIN & GIN-V & \approach \\ \hline
% \multicolumn{1}{|c|}{Accuracy} & 0.9096 & 0.9270 & 0.9302 & 0.9326 & 0.9277 & 0.9752 \\ \hline
% \end{tabular}}
% \end{table}

\begin{table}[!bt]
  \caption{Effectiveness (F1-Score) of \approach on all GNN configurations and Pre-training Strategies, for all downstream tasks. For each subcategory, the best 
%   average performance, maximum S.D and maximum variance 
results for each category are in \textbf{bold} text.}
  \vspace{-1em}
%   Details of Configuration Analysis of Pre-training Staretegies and GNN configurations for all downstream tasks}
 \scalebox{0.55}{
\begin{tabular}{|l|l|l|l|l|l|l|l|l|l|l|} %l|l|l|l|l|} %l|} 
\hline
 & \multicolumn{5}{c|}{\textbf{Method Name Prediction}} &  \multicolumn{5}{c|}{\textbf{Solution Classification}} 
%  &  \multicolumn{5}{c|}{\textbf{Code Clone Detection}}    
 \\ 
\textbf{GNN} & \textbf{No Pre-} & \multicolumn{3}{c|}{\textbf{Pre-training Strategies}}& \textbf{} & \textbf{No Pre-} & \multicolumn{3}{c|}{\textbf{Pre-training Strategies}} & \textbf{}  
% & \textbf{No Pre-} & \multicolumn{3}{c|}{\textbf{Pre-training Strategies}} & \textbf{}  
\\
& \textbf{training} & \textbf{Context} & \textbf{Node} & \textbf{VGAE} & \textbf{Average} & \textbf{training} & \textbf{Context} & \textbf{Node} & \textbf{VGAE} & \textbf{Average} 
% &\textbf{training} & \textbf{Context} & \textbf{Node} & \textbf{VGAE} & \textbf{Average}  
\\ \hline
GCN & 0.4494  &  0.5018 &  0.4859  & 0.5337 &  0.4930 & 0.9679  & 0.9710    & 0.9710  & 0.9751 & 0.9712 
% & 0.9995 & 0.9995    & 0.9993      & 0.9990    &  0.9991 
\\ 
GIN & 0.4347 & 0.4684   & 0.4037  & 0.5266  &  0.4584 & 0.9645 & 0.9711 & 0.9700    & 0.9710 & 0.9692 
% & 0.9984   & 0.9962   & 0.9974    & 0.9940    & 0.9965 
\\  
GraphSage & 0.3998 &  0.5006   & 0.4531  & 0.5412 & 0.4736 & 0.9675 & 0.9712   & 0.9721   & 0.9727 & 0.9709 
% & 0.9970  & 0.9993   & 0.9971   & 0.9986   & 0.9869 
\\  
GAT & 0.4246  & 0.5807     & 0.6194  & 0.5890 & 0.5534 & 0.9647  & 0.9746 &   0.9703   & 0.9735 & 0.9708 
% & 0.9732  & 0.9971     & 0.9956        & 0.9925   & 0.9869 
\\   
\hline
\textbf{Average} & 0.4271 & 0.5129 & 0.4905 & \textbf{0.5476} &  & 0.9662 & 0.9720 & 0.9718 & \textbf{0.9731} & 
% & 0.9920 & 0.9951 & \textbf{0.9973} & 0.9960 & 
\\   

\textbf{Variance} & 0.0003 & 0.0017 & \textbf{0.0064} & 0.0006 & & 2.2e-6 & \textbf{2.2e-6} & 7.1e-7 & 2.3e-6 & 
% & \textbf{0.0001} & 2.6e-5 & 1.7e-6 & 7.2e-6 & 
\\ 

\textbf{SD} & 0.0180 & \textbf{0.0413} & 0.0800 & 0.0244 &  & \textbf{0.0015} & \textbf{0.0015} & 0.0008 & \textbf{0.0015} & 
% & \textbf{0.0109} & 0.0051 & 0.0013 & 0.0027 & 
\\ 
\hline
\end{tabular}
}
\vspace{-2em}
\label{tab:config-analysis}
\end{table}

\begin{table}[!bt]
\centering
\caption{Ablation Study results showing the F1-Score of \approach. 
% on downstream tasks.
% , when only syntactic or semantic features are kept during the fine-tuning of \approach's pre-trained model. 
Best results 
% for each subcategory 
are \textbf{bold}.}
\vspace{-1em}
\label{tab:ablation}
\scalebox{0.61}{
\begin{tabular}{|l|l|cccc|cccc|}
\hline
% \multicolumn{2}{|l|}{\multirow{2}{*}{}} 
\textbf{Pre-training} & \textbf{Captured} & \multicolumn{4}{c|}{\textbf{Method Name Prediction}}   & \multicolumn{4}{c|}{\textbf{Solution Classification}}      \\ %\cline{3-10} 
% \multicolumn{2}{|l|}{}               
\textbf{Strategy} & \textbf{Feature}  & \textbf{GCN}    & \textbf{GIN}    & \textbf{GSAGE}  & \textbf{GAT}    & \textbf{GCN}    & \textbf{GIN}    & \textbf{GSAGE}  & \textbf{GAT}    \\ \hline
\multirow{2}{*}{\textbf{Context}}    & \textbf{semantic}    & \textbf{0.5454} & \textbf{0.4674} & \textbf{0.5038} & \textbf{0.6082} & \textbf{0.9698} & \textbf{0.9649} & \textbf{0.9682} & \textbf{0.9740} \\ 
                            & \textbf{syntactic}     & $0.4575$ & $0.4500$ & $0.4644$ & $0.4381$ & $0.9614$ & $0.9560$ & $0.9588$ & $0.9610$ \\ \hline
\multirow{2}{*}{\textbf{Node}}       & \textbf{semantic}    & \textbf{0.4843} & \textbf{0.4136} & \textbf{0.4404} & \textbf{0.5888} & \textbf{0.9738} & \textbf{0.9711} & \textbf{0.9696} & \textbf{0.9704} \\ 
                            & \textbf{syntactic}     & $0.3800$ & $0.3845$ & $0.3660$ & $0.3560$ & $0.9563$ & $0.9562$ & $0.9572$ & $0.9595$ \\ \hline
\multirow{2}{*}{\textbf{VGAE}}       & \textbf{semantic}    & \textbf{0.5988} & \textbf{0.4786} & $0.3675$ & \textbf{0.5464} & \textbf{0.9725} & \textbf{0.9663} & \textbf{0.9671} & \textbf{0.9711} \\  
                            & \textbf{syntactic}     & $0.3922$ & $0.4053$ & \textbf{0.3936} & $0.4058$ & $0.9711$ & $0.9659$ & $0.9626$ & $0.9705$ \\ \hline
 \multicolumn{2}{|c|}{\textbf{Best config.}}  &  \multicolumn{4}{c|}{\textbf{Semantic = 11/12}}  &  \multicolumn{4}{c|}{\textbf{Semantic = 12/12}}        \\ \hline               
\end{tabular}}
\vspace{-2em}
\end{table}

\smallskip\noindent
\textbf{RQ3 Further Analyses:}  
% \recheck{
The goal of this research question is to examine the impact of \textit{model pre-training} on improving \approach's effectiveness on downstream tasks. We also investigate if \approach effectively captures lexical and/or semantic program feature(s). % information. 
We employ \textit{probing analysis} to analyze if pre-trained \approach models learn the lexical and semantic features required for feature-specific tasks, i.e, that require capturing either or both features to be well-addressed. For instance, Task-4 is the concurrency classification task  requiring semantic features.
In addition, we conduct an \textit{ablation study} to investigate how the syntactic and semantic information captured by \approach influence its effectiveness on downstream tasks. Finally, we evaluate the \textit{sensitivity of our approach} to the selected \textit{GNN}.
% , and the impact of \textit{model pre-training} on improving \approach's effectiveness on downstream tasks.
% s}

% \smallskip
\noindent
% \textbf{RQ5 Effect of 
\textbf{\textit{Model Pre-training:} }
% \revise{ 
% Let us evaluate the impact of pre-training on the effectiveness of our approach. 
% In this evaluation, we examine the influence of model pre-training on the performance of our approach. 
% In particular, w
We examine if the three %self-learning 
pre-training strategies %of our approach 
improve the effectiveness of \approach on downstream tasks, using \recheck{two} downstream tasks and all three pre-training strategies (node, context and VGAE) 
%for this experiment 
% , 
% (\textit{see} \Cref{sec:approach}), 
% results are shown in 
% \todo{Figure X} and 
(\textit{see} \Cref{tab:config-analysis}). 
% provides the results on the effect of model pre-training on the performance of \approach. %details of 
% the GNN and 
% pre-training analysis of \approach. 
% }

% % \noindent
% % \textit{Pre-training Strategies: }
% \revise{
% In our evaluation, w
We found that \textit{model pre-training improves the effectiveness of \approach across all tasks}.  
% In particular, p
% In the best case, 
Pre-training improves %\approach's 
its effectiveness by up to 28\%, on average.  %For instance, 
For instance, consider %in the best case, 
% for 
model pre-training with VGAE strategy 
%improved the performance of \approach by \recheck{28}\% 
for method name prediction (\textit{see} \Cref{tab:config-analysis}). 
% , on average. 
% In the worst case, it improves performance for code clone detection by \recheck{}\%, on average.    
This result implies that %the importance of 
model pre-training improves the effectiveness of \approach on % addressing 
downstream SE tasks. 
% }.  
% }

\begin{result}
% \revise{
Model pre-training improves the effectiveness of \\ \approach (by up to 28\%, on average) across all tasks. 
% }
\end{result}

% \smallskip
\noindent
\textbf{\textit{Probing Analysis:} }
% \revise{
Let us examine if our pre-trained code embedding indeed encodes the desired lexical and semantic program features. %information.
To achieve this, we use the lexical embedding and semantic embedding from \approach's pre-training as inputs for probing. In this probing analysis, only the classifier is trainable and \approach is frozen and non-trainable. 
% For our probing analysis, w'
We use one MLP-layer classifier to evaluate these models on four tasks, % for our evaluation, 
Task-1 requires only lexical/syntactic information. However, Task-2 and Task-4 require only semantic information (program dependence). Finally, Task-3 subsumes tasks one and two, such that it requires both syntactic and semantic information.  
% \Cref{sec:experimental-setup} provides details of the probing experiments. %al setup. 
% about the setup for this experiment.
% }

Our evaluation results show that \textit{\approach's pre-trained code embedding mostly captures the desired lexical and semantic program features for all tested tasks, regardless of the pre-training strategy or GNN configuration}. 
\Cref{tab:probingTask} highlights the effectiveness of each frozen pre-trained model for each task, configuration and pre-training strategy. 
Notably, the frozen pre-trained model performed best for the desired embedding for each task in three-quarters (36/48=75\%) of all tested configurations. 
% Overall, the features encoded by the pre-trained model performed best for the task at hand in \recheck{75}\% of all tested configurations (36/48 configurations). 
As an example, for 
% instance, for 
tasks requiring semantic information (Task-2 and Task-4), 
% \approach's
our pre-trained model encoding only semantic information performed best 
% (better than capturing either or both information) 
for 88\% of all configurations (21/24 cases). 
% Meanwhile, for tasks requiring syntax only (Task-1), 
% % \approach's
% our pre-trained model encoding only syntactic information outperformed all other models for most (8/12=67\%) configurations. % ( cases).
% Besides, for this task (Task-1), syntactic code embedding models performed better than the semantic-only models, as well as the syntactic and semantic (syn+sem) models. 
% Finally, \approach's pre-trained model encoding both syntactic and semantic information (Task-3) outperformed models capturing only either syntactic or semantic information in most (7/12=58\%) cases. %Overall, t
This result demonstrates that \approach effectively encodes 
either or both 
syntactic and semantic features, this % This 
is evidenced by the effectiveness of models encoding %each or both 
desired 
feature(s) for feature-specific tasks. % (Tasks one to four). % requiring such features. 
% \revise{}

% \recheck{
% For Task-1, \tabref{tab:probingTask} 
% shows that the token representation achieves better performance than the structure representation. 
% As expected, in Task-2, the result is reverse as shown in \tabref{tab:probingTask}:
% Structure representation is better than the token representation.
% %The difference is due to the different characteristics (\secref{sec:probdatasetconstruct}) between LeetCode-10 and M-LeetCode. 
% %The probing Task-3 is a more difficult task than the first two, which subsumes Task-1 and Task-2. 
% For Task-3 results in \tabref{tab:probingTask}, in some cases, the structure representation obtains higher accuracy 
% than the token representation, and the conclusion is flipped in some other scenarios. 
% Task-4 Concurrency Detection gets similar results with Task-2, recalling that concurrency detection need more structure information than the token information. 
% Therefore, pretrained \approach encodes the expected token and structure information of the program.
% }

\begin{result}
% \revise{
\approach effectively encodes the syntactic and/or semantic features, feature-specific models 
% encoding 
% requiring 
% each or both of these 
% desired features 
performed best in 75\% of cases.
% tested cases across all feature-specific tasks. 
% this is evident by the ....  configuration  three-quarters  in  
% }
\end{result}

\noindent
\textbf{\textit{Ablation Study:} }
% \revise{
We investigate the impact of syntactic/lexical embedding and semantic/dependence embedding on addressing downstream tasks. %, we used two tasks, namely 
Using method name prediction and solution classification, we examine how removing lexical embedding or dependence embedding during the fine-tuning of \approach's pre-trained model impacts the effectiveness of the approach. 
% This experiment shows the importance of lexical and dependence features on the effectiveness of \approach.% for these downstream tasks. 
% }

% \revise{
Our results show that \textit{\approach's 
% it is important for \approach to 
dependence %or semantic 
embedding 
% semantic information 
is important to %be effective in 
effectively address our downstream SE tasks}. \Cref{tab:ablation} presents the 
%rovides details on the results of our 
ablation study results. In particular, results show that models fine-tuned with only semantic information outperformed those fine-tuned with syntactic features in almost all (23/24 = 96\% of) cases. This result demonstrates the effectiveness of dependence embedding 
% e information in
% features in the code embedding  
% to address 
in addressing %these 
downstream SE tasks. 
% }

\begin{result}
% \revise{
%  \approach 
Results show that dependence/semantic embedding is vital to the effectiveness of \approach on downstream SE tasks.
%  it is vital to captur/e 
% }
\end{result}

% \recheck{
% We investigate the effects of each representation in our approach in two downstream tasks, Java-Small and Java250, instead of the default concatenation of both of the representations, to see how does it affect the performance if we remove the tokenrepresentation or the structure representation when finetun-ing the pretrained models. \tabref{tab:ablation} demonstrate the ablation result. Compared with Table \ref{tab:config-analysis}, we can see \approach performance decreases in the ablation experiment \tabref{tab:ablation}. Meanwhile, structure representation gets the better performance than the token representation as shown in \tabref{tab:ablation}. Structure representation reflects the hybrid role of the node \cite{rossi2014role} in the structure and it also contains some token information. However, the structure representation loses some token information with the objective learning program structure, and the token representation can be complementary with the loss of the structure representation.Overall, it can improve \approach performance performance by considering both of the token and structure representation and removing either of them hurts \approach performance.
% }

%\todo{we need some charts and figures.\com{Has the todo been done?}}

% \smallskip
\noindent
\textbf{\textit{GNN Sensitivity:} }
% \revise{
This experiment evaluates the sensitivity of our approach to the choice of GNN. 
% In this experiment, we employ both %all three 
% downstream tasks, and four GNN configurations, namely GAT, GIN, GCN and GraphSAGE  (\textit{see} \Cref{sec:approach}). %\todo{Figure X} and 
\Cref{tab:config-analysis} provides details of the GNN sensitivity analysis, tasks and GNN configurations. To evaluate this, % end, 
we compute %We demonstrate this by computing 
the \textit{variance} and \textit{standard deviation (SD)} of the effectiveness of \approach when employing different GNNs. % of \approach.
% }

% \noindent
% \textit{GNN configurations: }

% \revise{
Our evaluation results show that \textit{\approach is stable, it is not highly sensitive to the choice of GNN}. \Cref{tab:config-analysis} shows the details of the SD and variance of our approach for each GNN configuration. Across all tasks, the variance and SD of the 
% \performance of 
\approach is  mostly low, it is maximum 0.0064 and 0.0413, respectively.
% (\textit{see} \Cref{tab:config-analysis}). 
% % For instance, the variance and standard deviation for GCN is between \todo{X and X}. This is a 
% This result implies that \approach is very stable. 
% , i.e., it is not highly sensitive to the choice of GNN. 
% }

\begin{result}
% \revise{
\approach is stable across GNN configurations, 
the %its 
% (maximum) 
variance and SD of its effectiveness
% standard deviation 
are very low for all configurations. 
% }
\end{result}

% \begin{table*}[bt]
%   \caption{Performance of GNN configurations and Pre-training Strategies for all downstream tasks}
%  \scalebox{0.8}{
% \begin{tabular}{|l|l|l|l|l|l|l|l|l|l|l|l|l|l|} %l|} 
% \hline
%  & \multicolumn{4}{c|}{\textbf{Method Name Prediction}} &  \multicolumn{4}{c|}{\textbf{Solution Classification}} &  \multicolumn{4}{c|}{\textbf{Code Clone Detection}} &   \\ 
% \textbf{GNN} & \textbf{No Pre-} & \multicolumn{3}{c|}{\textbf{Pre-training Strategies}} & \textbf{No Pre-} & \multicolumn{3}{c|}{\textbf{Pre-training Strategies}} & \textbf{No Pre-} & \multicolumn{3}{c|}{\textbf{Pre-training Strategies}} & \textbf{Average} \\
% & \textbf{training} & \textbf{Context} & \textbf{Node} & \textbf{VGAE}  & \textbf{training} & \textbf{Context} & \textbf{Node} & \textbf{VGAE}& \textbf{training} & \textbf{Context} & \textbf{Node} & \textbf{VGAE} & \\ \hline
% GCN & & & & & & & & & & & & & \\ 
% GIN & & & & & & & & & & & & & \\  
% GraphSage & & & & & & & & & & & & & \\  
% GAT & & & & & & & & & & & & & \\   
% \hline
% \textbf{Average} & & & & & & & & & & & & & \\   
% \hline

% \end{tabular}
% }
% \label{tab:config-analysis}
% \end{table*}

%\input{discussions}
% \section{Discussions and Future Outlook}
% \label{sec:discussions}

%\input{threats-to-validity}

\section{Threats to Validity}
\label{sec:threats-to-validity}

\noindent
\textit{External Validity:} 
% \revise{
This refers to the generalizability of our approach and results, especially beyond our data sets, tasks and models. For instance, 
% our approach may not generalize to other programming languages other than Java. Besides, 
there is a threat that \approach does not generalize to other (SE) tasks and other Java programs. 
% We have also evaluated our approach on three downstream tasks, three Java subject programs and a limited training data set. 
To mitigate this threat, we have evaluated \approach using mature Java programs with varying sizes and complexity (\textit{see} \Cref{tab:subject_programs}), as well as downstream tasks with varying complexities and requirements. 
% We have also employed a considerable diverse set of programs for pre-training, our \texttt{Jimple-Graph} dataset is from 46 different domains (\textit{see} \Cref{sec:experimental-setup}). 
% We are also confident in our evaluation because our subject programs are large (\todo{X} KLOC), complex (\todo{X} methods) and mature (\todo{X} years of maturity). 
% In addition, %several of 
% our selected downstream SE tasks have been extensively studied by the research community for several decades (e.g., code clone detection~\cite{sheneamer2016survey}).    
% }

\noindent
\textit{Internal Validity:} 
% \revise{
This threat refers to the correctness of our implementation, %in particular, 
if we have correctly represented lexical and semantic features in our code embedding.
% and model (pre-)training. 
We mitigate this threat by evaluating the %\recheck{
validity of our implementation with probing analysis and ablation studies 
% to ensure our models work as expected 
(\textit{see} \Cref{sec:results}). We have also compared \approach to \recheck{7 baselines %approaches 
using four (4) major downstream tasks}. In addition, we have conducted further analysis to test our implementation using different pre-training strategies and GNN configurations. 
% Hence, we are confident about the correctness of our approach. 
We also provide our implementation, (pre-trained) models and experimental data for scrutiny, replication and reuse.  
% }

\noindent
\textit{Construct Validity:} 
% \revise{
This is the threat posed by our design/implementation choices and their implications on our findings. Notably, our choice of intermediate code representation (i.e., Jimple) instead of source code implies that our approach lacks natural language text (such as code comments) in the (pre-)training dataset. 
% Such information 
% As an example, natural language text is vital 
% may be necessary 
% for some downstream SE tasks such as comment generation and code summarizing. 
% 
Indeed, \approach would not capture this information as it is. However, it is possible to extend 
\approach
% it to work at the source code level 
to also capture natural language text. %For instance, t
This can be achieved by performing lexical and program dependence analysis at the source code level. 
% similar to CodeBERT~\cite{feng2020codebert} and GraphCodeBERT~\cite{guo2020graphcodebert}. 
% }

% \todo{Intermediate representation Jimple vs. source code}
% \todo{Java only, }
% \todo{limited SE/downstream tasks}
% \todo{limited datasets, }

%\input{conclusion}
\section{Conclusion}
\label{sec:conclusion}
% \revise{
In this paper, we have proposed \approach, a novel and generic code embedding approach that captures both syntactic and semantic program features. % of the program. 
 We have evaluated it in comparison to 
 % that it is more effective than 
 the state-of-the-art generic code embedding approaches, as well as specialised, task-specific learning based applications. Using \recheck{seven (7) baselines and four (4) major downstream SE tasks}, we show that \approach %has been demonstrated to be 
is stable and effectively applicable to several downstream SE tasks, e.g., patch classification and 
% , code clone detection and 
solution classification.
 Moreover, we show that it indeed captures both lexical and 
%  semantic
 dependency features, and we demonstrate the importance of generically embedding both features to solve downstream SE tasks.  
We also 
provide our experimental %models and 
code for replication and reuse: 
% \parskip

 \begin{center}
 \vspace{-0.2mm}
     \textbf{\url{https://github.com/graphcode2vec/graphcode2vec}}
 \end{center}
% }

%\begin{acks}

%\end{acks}

%\newpage

\balance
\bibliographystyle{ACM-Reference-Format}
\bibliography{reference}

\end{document}